\PassOptionsToPackage{pdfpagelabels=false}{hyperref} 
\documentclass[useAMS,usenatbib]{mnras}
\pdfoutput=1
\usepackage[pdftex]{graphicx}
\usepackage{newtxtext,newtxmath}
\usepackage[T1]{fontenc}
\usepackage{xspace}
\usepackage{bm}
\newcommand{\msun}{\,M$_{\odot}$\xspace}
\newcommand{\msunyr}{\,M$_{\odot}$\,yr$^{-1}$\xspace}
\newcommand{\sbunits}{\,erg\,s$^{-1}$\,kpc$^{-2}$\xspace}
\newcommand{\oviir}{\,\ion{O}{VII}{\small (r)}\xspace}

\usepackage{xcolor}

\volume{{\rm submitted}}

\title[OVII X-ray resonant scattering]{Resonant scattering of the OVII X-ray emission line in the \\circumgalactic medium of TNG50 galaxies}
\author[D. Nelson et al.]{Dylan Nelson$^{1}$\thanks{E-mail: dnelson@uni-heidelberg.de}, Chris Byrohl$^{1}$, Anna Ogorzalek$^{2,3}$, Maxim Markevitch$^{2}$, Ildar Khabibullin$^{4,5}$, \newauthor
Eugene Churazov$^{5}$, Irina Zhuravleva$^{6}$, Akos Bogdan$^{7}$, Priyanka Chakraborty$^{7}$, Caroline Kilbourne$^{2}$, \newauthor Ralph Kraft$^{7}$, Annalisa Pillepich$^{8}$, Arnab Sarkar$^{9}$, Gerrit Schellenberger$^{7}$, Yuanyuan Su$^{10}$, \newauthor Nhut Truong$^{8}$, Stephan Vladutescu-Zopp$^{4}$, Nastasha Wijers$^{11}$
\\\\
$^{1}$ Universit\"{a}t Heidelberg, Zentrum f\"{u}r Astronomie, Institut f\"{u}r Theoretische Astrophysik, Albert-Ueberle-Str. 2, 69120 Heidelberg, Germany\\
$^{2}$ NASA Goddard Space Flight Center, Greenbelt, MD 20771, USA\\
$^{3}$ Department of Astronomy, University of Maryland, College Park MD 20742-2421, USA\\
$^{4}$ Universit\"{a}ts-Sternwarte, Fakult\"{a}t f\"{u}r Physik, Ludwig-Maximilians-Universit\"{a}t M\"{u}nchen, Scheinerstr. 1, 81679 M\"{u}nchen, Germany\\
$^{5}$ Max-Planck-Institut f\"{u}r Astrophysik, Karl-Schwarzschild-Str. 1, 85741 Garching, Germany\\
$^{6}$ Department of Astronomy and Astrophysics, The University of Chicago, Chicago, IL 60637, USA\\
$^{7}$ Center for Astrophysics | Harvard \& Smithsonian, 60 Garden Street, Cambridge, MA, 02138, USA\\
$^{8}$ Max-Planck-Institut f\"{u}r Astronomie, K\"{o}nigstuhl 17, 69117 Heidelberg, Germany\\
$^{9}$ Kavli Institute for Astrophysics and Space Research, Massachusetts Institute of Technology, 77 Massachusetts Ave, Cambridge, MA\\
$^{10}$ University of Kentucky, 505 Rose Street, Lexington, KY 40506, USA\\
$^{11}$ CIERA and Department of Physics and Astronomy, Northwestern University, 1800 Sherman Ave, Evanston, IL 60201, USA
\vspace{-2em}
}

\begin{document}

\maketitle

\begin{abstract}
We study the impact of resonantly scattered X-ray line emission on the observability of the hot circumgalactic medium (CGM) of galaxies. We apply a Monte Carlo radiative transfer post-processing analysis to the high-resolution TNG50 cosmological magnetohydrodynamical galaxy formation simulation. This allows us to model the resonant scattering of \oviir X-ray photons within the complex, multi-phase, multi-scale CGM. The resonant transition of the \ion{O}{VII} He-like triplet is one of the brightest, and most promising, X-ray emission lines for detecting the hot CGM and measuring its physical properties. We focus on galaxies with stellar masses $10.0<\log{(M_\star/\rm{M_\odot})}<11.0$ at $z\simeq0$. After constructing a model for \oviir emission from the central galaxy as well as from CGM gas, we forward model these intrinsic photons to derive observable surface brightness maps. We find that scattering significantly boosts the observable \oviir surface brightness of the extended and diffuse CGM. This enhancement can be large -- an order of magnitude \textit{on average} at a distance of 200 projected kpc for high-mass $M_\star=10^{10.7}$\msun galaxies. The enhancement is larger for lower mass galaxies, and can even reach a factor of 100, across the extended CGM. Galaxies with higher star formation rates, AGN luminosities, and central \oviir luminosities all have larger scattering enhancements, at fixed stellar mass. Our results suggest that next-generation X-ray spectroscopic missions including XRISM, LEM, ATHENA, and HUBS -- which aim to detect the hot CGM in emission -- could specifically target halos with significant enhancements due to resonant scattering.
\end{abstract}



\section{Introduction}

The hot phase of the circumgalactic medium surrounding galaxies is the dominant baryonic reservoir of dark matter halos. Its thermodynamical and cooling properties are largely responsible for the balance between cosmic gas accretion on to, and feedback-driven outflows from, the central galaxy. The hot circumgalactic medium (CGM) therefore regulates the baryon cycle, as well as the process of galaxy evolution \citep[reviewed in][]{donahue22}.


One way to observationally detect and characterize the hot CGM is through its X-ray emission. X-rays are a key tool to probe massive $M_{\rm halo} > 10^{14}$\msun clusters \citep{pratt09,vikhlinin09,mcdonald13}, as well as $10^{13} < M_{\rm halo}/\rm{M}_\odot < 10^{14}$ groups \citep{lovisari15,eckert21}. While individual detections of the hot CGM are challenging for Milky Way-mass galaxies with $M_{\rm halo} \sim 10^{12} - 10^{12.5}$\msun \citep{anderson11,bogdan13,li13}, stacking with ROSAT \citep{anderson15} and recently eROSITA \citep{comparat22,chadayammuri22} can access this regime. However, measuring the physical properties of the CGM, such as temperature, abundance, density, and velocity is difficult at CCD spectral resolution \citep{kraft22}. This motivates the need for future X-ray imaging spectrographs.

At Milky Way through cluster mass scales, direct imprints of astrophysical feedback processes from galaxies are expected in the physical properties, and observables, of the hot CGM \citep{nelson18b,truong20,oppenheimer20,truong21}. These gaseous halos contain a relatively small fraction of the baryons expected within $\Lambda$CDM for a given dark matter halo mass \citep{anderson10}, the remainder having been ejected to the larger scale of the closure radius \citep{ayromlou23}.


High-mass $T_{\rm vir} \gtrsim 10^7$\,K clusters emit primarily via free-free bremsstrahlung due to their high temperatures, as reflected in cluster X-ray scaling relations \citep{robson20,pop22}. In contrast, lower mass halos are dominated almost exclusively by emission from metal lines. The observational detectability of this diffuse gas has been theoretically studied using modern cosmological hydrodynamical simulations (\citet{wijers22}, \textcolor{blue}{Schellenberger et al. in prep}). Mapping its spatial variation with high angular resolution imaging, and its spectral variation with high energy resolution X-ray spectroscopy, offers powerful probes of the physical properties of the CGM as well as the underlying connections to galaxy feedback processes (\textcolor{blue}{Truong et al. in prep}, \textcolor{blue}{ZuHone et al. in prep}). Forward modeling enables quantitative theoretical predictions for observable X-ray photons \citep{zuhone22} and links those observables back to the underlying physical models \citep{truong21b}.


However, a significant complication exists for the detection and interpretation of X-rays from the CGM. While the hot circumgalactic medium has a small optical depth for continuum X-ray photons, this is not necessarily true for the emission lines of resonant transitions of highly ionized metals \citep{churazov04}. For example, in massive clusters the optical depths of iron lines including Fe XXV and Fe XXIV can range from of order a few to $\sim 10$ from the center \citep{gilfanov87}. In the absence of scattering, the radial surface brightness profile of an emission line will follow that of the continuum. For a constant total luminosity, resonant scattering will however redistribute emission into the outskirts, lowering the central surface brightness while boosting it at larger distance \citep{churazov10,hitomi18b}. This effect is analogous to that in the resonant Lyman-alpha line of hydrogen, where scattering significantly flattens the intrinsic surface brightness profile \citep{byrohl21,lujanniemeyer22}. The scattering of X-ray photons from the diffuse cosmic background, in the diffuse intergalactic medium, can also impact the observability of the IGM in emission \citep{churazov01,khabibullin19}.


For Milky Way to group mass halos with $10^{12} - 10^{13}$\msun total mass, lines of ionized \ion{O}{VII} and \ion{O}{VIII} are particularly important, as these ions are the dominant ionization states of oxygen \citep{nelson18b}. We focus here on \ion{O}{VII}, which is He-like with two bound electrons remaining \citep{chakraborty21}. The atomic physics of the electronic transitions of He-like ions are well understood \citep{porquet10}. The $n=2$ excited state is split into four levels, of which three decay\footnote{The fourth excited state can only de-excite via a two-photon channel, producing continuum-like emission.} to the $n=1$ ground state, leading to a triplet of emission lines. These are: the $^{1}\rm{P_1}$ resonant line (labeled r), the $^{3}\rm{P_i}$ intercombination lines (i), and the $^{3}\rm{S_1}$ forbidden line (f). In the absence of scattering effects, the triplet component ratios are sensitive to the physical state of the plasma, making it a potential diagnostic of gas density and temperature \citep{ezoe21}.


The resonant \oviir line at $21.602$\AA\ has a transition energy of $0.574\,$keV in the soft X-ray band. Its resonant nature and considerable optical depth implies that \oviir photons scatter off of intervening gas before reaching the observer, modulating observables including surface brightness profiles, spectral shapes, and polarization \citep{churazov10}. For example, we can observe radiation emitted in the past by a bright central source such as an active supermassive black hole (AGN), as it scatters off CGM gas. Such an echo of previous AGN activity would constrain the past luminosity, activity, and therefore duty cycles of AGN \citep{sunyaev93,sazonov02}. Even more intriguing, because \oviir photons sample the scattering medium, they are sensitive to the full three-dimensional velocity field of the CGM. The resulting spectral profiles therefore simultaneously encode information on microturbulent (small-scale) gas motions as well as large-scale motion and bulk flows \citep{sanders06,zhuravleva11}. Unlike most other extragalactic velocity observables, we can therefore probe gas motion in all directions, either through spectral distortions or polarization \citep{zhuravleva10}, as opposed to the line-of-sight component alone.


Accessing the information content of X-ray emission lines such as the \ion{O}{VII} triplet and \oviir in particular requires high spectral resolution. To date, instrumentation has restricted the observational possibilities: the RGS onboard XMM-Newton has been used to measure the characteristic turbulent velocity and impact of resonant scattering with the \ion{Fe}{XVII} line \citep{werner09}. Turbulent velocities have also been inferred via \ion{Fe}{XVII} for samples of nearby elliptical galaxies \citep{deplaa12,ogorzalek17}. However, the modern era of high-resolution X-ray spectroscopy began with the observations of turbulence in the Perseus cluster core by Hitomi \citep{hitomi18}, which also detected resonant scattering of the FeXXV He$\alpha$ line \citep{hitomi18b}.

The impending launch of XRISM \citep{xrism20} promises to provide detailed observations of X-ray emitting gas in galaxy clusters, constraining the physics of turbulence, enrichment, and mixing \citep{simionescu19,simionescu20}. Ambitious future mission concepts will, for the first time, probe the physics of the hot CGM down to Milky Way-mass halo scales. The X-IFU instrument onboard ATHENA \citep{nandra13} will offer an unparalleled view on the physics of the energetic Universe and the hot X-ray emitting gas in and around dark matter halos \citep{barret13,kaastra13}.

The Line Emission Mapper \citep[LEM;][]{kraft22} NASA Probe concept is an imaging spectrometer based on $\sim 120 \times 120$ pixel TES microcalorimeter array with $\sim$eV spectral resolution, a 30' field of view, and a 15" angular resolution, focusing its effective area entirely at the 0.2-2 keV soft X-ray range. With $\sim$\,1Ms pointings on a large number of nearby Milky Way-mass halos, it will map the distribution and kinematics of the hot CGM. One of its principal targets will be the forbidden line of \ion{O}{VII}, redshifted away from the Milky Way foreground at the same wavelength. As we argue herein, the surface brightness dimming effect of moving to slightly higher redshift targets may be offset, or even reversed, due to resonant scattering enhancement of \oviir, making it a compelling target.


In this work we combine the TNG50 cosmological magnetohydrodynamical galaxy formation simulation with a Monte Carlo radiative transfer method. We construct an emission model for the resonant \oviir line of O$^{6+}$ oxygen ions, and focus on galaxies with stellar masses $10.0 < \log{(M_\star / \rm{M_\odot})} < 11.0$ at $z \simeq 0$. We forward model these intrinsic photons via resonant scattering into observable surface brightness maps around individual galaxies. Capturing the multi-phase, multi-scale CGM, this enables us to study the impact of X-ray resonant scattering on the observability of the hot CGM.

The paper is organized as follows: Section \ref{sec_methods} outlines the methodology, including the radiative transfer approach, Section \ref{sec_results} presents our main results, and Section \ref{sec_conclusions} summarizes our findings.


\section{Methods} \label{sec_methods}

\subsection{The TNG50 Simulation} \label{sec_sims}

Given its combination of high numerical resolution and large volume, the TNG50 cosmological magnetohydrodynamical simulation \citep{pillepich19, nelson19b} is ideal for the current study. The TNG50 volume spans $\sim$(50 cMpc)$^3$, with a baryonic (gas/star) mass resolution of $\sim 8 \times 10^4$ M$_\odot$. It is the third volume of the IllustrisTNG project \citep[TNG hereafter;][]{springel18, naiman18, pillepich18b, marinacci18, nelson18a}, a suite of cosmological simulations which are the successor of the original Illustris simulation \citep{vog14a, vog14b, genel14, sijacki15}. 

The TNG simulations are run with the \textsc{Arepo} moving-mesh code \citep{spr10}, and all adopt the same `TNG model' for galaxy formation physics \citep{weinberger17, pillepich18a}, including the effects of magnetic fields using an ideal magnetohydrodynamics, divergence cleaning method \citep{pakmor11,pakmor14}. Self-gravity is solved with the Tree-PM scheme, while the fluid dynamics is based on a Godunov-like finite-volume scheme, using an unstructured, moving, Voronoi tessellation to discretize space.

The TNG simulations include a broad and well-tested physical model for the most important processes driving the formation and evolution of galaxies. In brief: (i) gas microphysics and radiative processes, including primordial/metal-line cooling and heating from the background radiation field i.e. UVB, (ii) star formation in the dense interstellar medium, based on a density-threshold model, (iii) stellar population evolution, chemical enrichment, and metal return, following supernovae type Ia, II, as well as AGB stars, and individually tracking nine elements: H, He, C, N, O, Ne, Mg, Si, and Fe, (iv) galactic-scale outflows driven by stellar i.e. supernovae feedback \citep[see][]{pillepich18a}, (v) supermassive black holes: their seeding/formation, coalescence/mergers, and growth via gas accretion, (vi) and blackhole feedback, operating in a time-continuous thermal mode at high accretion rates, and a high-velocity, time stochastic, kinetic wind mode at low accretion rates \citep[see][]{weinberger17}.

Consistent with the TNG simulations, we adopt a \cite{planck2015_xiii} cosmology with $\Omega_{\Lambda,0}=0.6911$, $\Omega_{m,0}=0.3089$, $\Omega_{b,0}=0.0486$, $\sigma_8=0.8159$, $n_s=0.9667$ and $h=0.6774$. 

\subsection{\oviir Emission Model}

Our modeling of the emissivity of diffuse gas follows directly the methodology of \citet{nelson21} for \ion{Mg}{II} emission, which we summarize briefly here. We compute the emission from the \oviir line using \textsc{Cloudy} \citep[][v17.00]{ferland17}, including both collisional and photo-ionization processes, and assuming ionization equilibrium given a UV + X-ray background \citep[][FG11 update, making this choice fully self-consistent with the simulations]{fg09}. We account for self-shielding with a frequency dependent attenuation of the UVB at high densities \citep[following][]{bird14,rahmati13}. We use \textsc{Cloudy} in its single-zone mode and iterate to equilibrium, running in the constant temperature mode, with no induced processes \citep[following][]{wiersma09} and assuming the solar abundances of \cite{grevesse10}.

Emissivities $\epsilon_{\rm V}(n_{\rm H},T,z,Z)$ are tabulated over a 4D grid of hydrogen number density, temperature, redshift, and metallicity. To derive the emission from each gas cell in the simulation, we interpolate in this table, taking into account the actual gas-phase oxygen abundance, as tracked directly in the simulation, following its production in stars and subsequent return and mixing \citep{pillepich18a}.\footnote{We note that the vast majority of oxygen is produced promptly via the SNII channel, for which TNG adopts a combined stellar yield model from \citet{kobayashi06} and \citet{portinari98}. At solar metallicity, within a Hubble time, $\sim$\,1.8\% per unit stellar mass formed returns to the ISM as oxygen. AGB stars contribute 1 dex less, and SNIa 1 dex less still.}

Dense, star-forming gas cells in the TNG model adopt a two-phase pressurization model \citep{spr03}. In contrast to our previous emission models, which simply assumed that the total mass of such cells exists at their cold-phase temperature of 1000\,K, we here develop a new two-phase emission model for star-forming ISM gas. Specifically, we calculate the mass fraction of the hot and cold components, as well as the temperature of the hot phase, which are all density dependent \citep{spr03}. The mass fraction of the hot phase is small -- roughly 10\% at the star-formation threshold density of $n_{\rm H} \simeq 0.1 \rm{cm}^{-3}$, and decreases to $\lesssim 1$\% with increasing density \citep[see e.g.][]{stevens19}. The temperature of the hot phase increases asymptotically from $\sim 10^5$\,K at $n_{\rm H} \simeq 0.1 \rm{cm}^{-3}$ to $\sim 10^8$\,K at the highest densities. As the mean density of star-forming gas in high-resolution TNG model simulations is $\bar{n}_{\rm H,SF} \simeq 1 \rm{cm}^{-3}$, the corresponding hot-phase temperature of $\sim 10^6$\,K can efficiently host high abundances of \ion{O}{VII} \citep{nelson18b}.

To model the \oviir emission from multi-phase ISM gas we derive the hot-phase density from the respective mass fraction, and adopt its corresponding density-dependent temperature for the respective \textsc{Cloudy} calculation. This is an approximate model for the emission from the hot interstellar medium of galaxies, and is primarily meant to give a non-vanishing, density dependent emissivity which can then be scaled as needed. In particular, we introduce a `boost parameter', $b=10^{-4}$ being our fiducial value, which we treat as a free parameter to scale up or down this emission component. A value of $b=0$ corresponds to zero emission from star-forming gas, while a value of $b=1$ corresponds to the naive application of the described model (as we show below, $b=1$ produces unphysically high luminosities in galaxies). As the hot ISM is within the central galaxy itself, i.e. at small radii $\lesssim 10$\,kpc, our $b$-parameter also allows us to experiment with the impact of a central, bright source, which in reality could be the galaxy itself, or a luminous AGN. A low or zero $b$ value therefore also corresponds to negligible or no \oviir emission from the AGN itself.

Our model makes several assumptions. First, individual gas cells i.e. at the simulation resolution limit have constant density and temperature: no emission arises from gas structure inhomogeneity on physical scales $\lesssim $100 pc, which are unresolved in TNG50. Second, we do not include any stellar continuum emission at the frequency of the \oviir line. Third, we omit the impact of radiation from local stellar/AGN sources, which could change oxygen ions fractions near galaxies \citep{suresh17,oppenheimer18a}.

\subsection{Resonant scattering radiative transfer}

\begin{figure*}
\centering
\includegraphics[angle=0,width=1.0\textwidth]{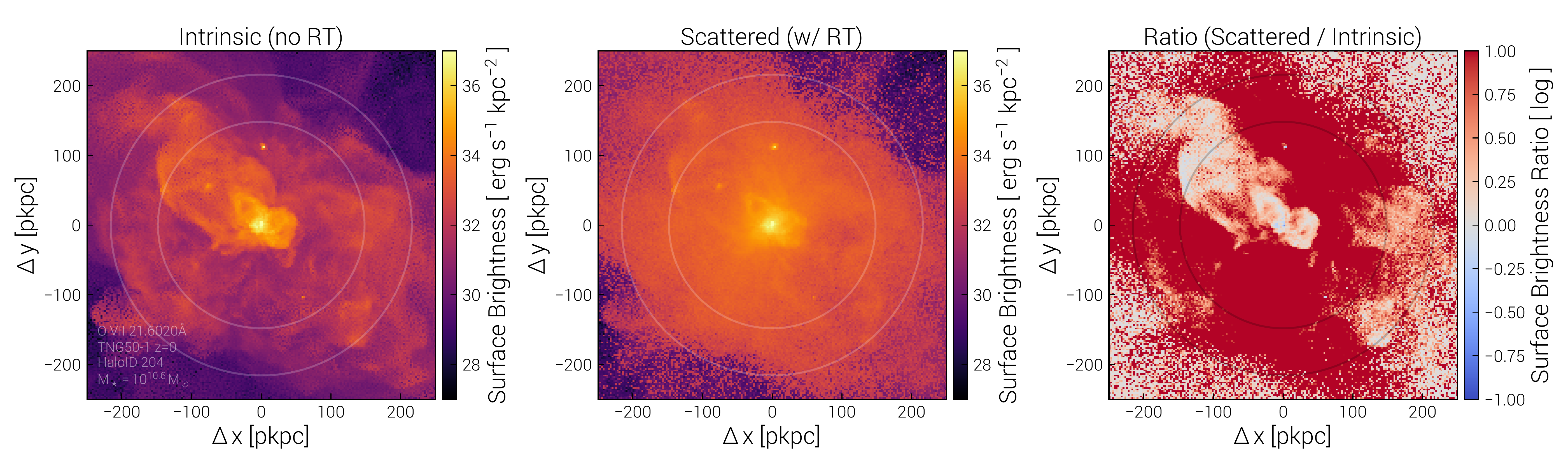}
\caption{The impact of resonant scattering on the \oviir surface brightness from the circumgalactic medium of a single TNG50 galaxy. The left panel shows intrinsic \oviir emission, without radiative transfer effects i.e. neglecting scattering. In contrast, the middle panel shows the scattered surface brightness map (same color scale), after forward modeling photons with radiative transfer. Only the middle panel is an observable. The right panel shows the log$_{10}$ ratio of the two: white regions have unchanged surface brightness, blue regions have suppressed emission after scattering, and red regions have enhanced emission after scattering. This prototypical example is Milky Way-like, with a stellar mass $M_\star = 10^{10.6}$\msun (halo ID 204). The inner and outer circles mark $R_{\rm 500c}$ and $R_{\rm 200c}$, respectively. Resonant scattering redistributes emission from the bright central region, as well as from bright outflow features (upper left) into the extended, volume filling CGM. The observable \oviir surface brightness is enhanced by more than an order of magnitude across most of the halo, out to the virial radius.
 \label{fig_single_maps}}
\end{figure*}

Unlike \citet{nelson21}, we do not assume that photons propagate through an optically thin medium. Instead, we introduce a full radiative transfer (RT) treatment of the resonant scattering process. To do so, we extend the Monte Carlo RT method of \cite{byrohl21,byrohl22}, previously used to study Lyman-alpha, to resonant metal lines, including \oviir. This transition occurs at $\lambda = 21.602$\AA\ and we adopt an oscillator strength of $f = 0.696$ and emission coefficient $A = 3.32 \times 10^{12}$\,s$^{-1}$ \citep{verner96}.

While the multi-scattering physics is similar to the Lyman-$\alpha$ line of hydrogen \citep{prochaska11a}, the optical depths involved are much lower. In particular, resonantly trapped \ion{O}{VII} photons require only $\mathcal{O}(1 - 10)$ scatterings to escape a CGM environment, similar to the situation for \ion{Fe}{XXV} in clusters, and \ion{Fe}{XVII} in groups. This allows us to disable the numerical acceleration schemes, including core skipping, used in \cite{byrohl21} to speed up the frequency diffusion process. The resulting RT of \oviir is relatively straightforward in comparison to Lyman-alpha.

\oviir photons are emitted at the velocity of each moving gas cell, meaning that bulk velocities as well as (resolved) turbulent velocities are fully taken into account, for the emitting gas as well as the scattering gas. We do not include an additional turbulent velocity component to account for unresolved turbulence, i.e. on scales $\lesssim 100$\,pc, which would introduce an ad hoc assumption. Photons have an initial wavelength distribution with respect to the line-center frequency corresponding to a thermal Gaussian, given the temperature of the gas. Photons are emitted with a random initial direction, and we adopt an isotropic scattering phase function, a simplification which is expected to have negligible impact on our results. After each scattering, we calculate a luminosity contribution which escapes towards a defined observer viewing direction~\citep[the peeling-off algorithm;][]{Whitney11}. During the RT, \oviir photons only scatter (via absorption followed by rapid re-emission), they are not destroyed, and do not alter the gas state (its temperature or ionization). Our RT method can ray-trace through a variety of geometrical discretizations of the underlying gas, including unstructured Voronoi tessellations, which we use here for full consistency with TNG50.

\subsection{Galaxy sample}

We process each halo individually, propagating photons out to twice the virial radius, and neglecting any IGM-scale scattering effects. Our sample is drawn from TNG50 at $z=0$. We select central galaxies with $10.0 < \log{(M_\star / \rm{M_\odot})} < 11.0$. There are a total of 479 such galaxies, and we randomly sub-select up to 30 per 0.1 dex bin of $M_\star$. Our final sample contains 298 galaxies.

We do not apply any explicit observational realism effects, such as noise, or a finite angular resolution/point spread function. Our conclusions are thus general and apply broadly to future X-ray spectroscopic instruments including XRISM, LEM, ATHENA, and HUBS.


\section{Results} \label{sec_results}

We begin with a demonstration of the impact of resonant scattering for a single TNG50 galaxy. Figure \ref{fig_single_maps} shows the predicted \oviir surface brightness from the scale of the circumgalactic medium (CGM) of this system. We contrast two cases: intrinsic photons, where we neglect radiative transfer effects and any scattering (left panel), and scattered photons, where we process this emission with our radiative transfer method (middle panel). We find a striking visual difference. In particular, the scattered surface brightness map -- that is, the observable emission -- is significantly brighter throughout large areas of the halo. \oviir photons originating from the central galaxy scatter off gas throughout the CGM before last scattering towards the observer, significantly boosting its surface brightness. The circumgalactic medium in \oviir is largely illuminated by light which is originally emitted at its very center.

Each map is shown 250 kpc across, and through an equal projection depth along the line of sight direction. The viewing direction i.e. galaxy orientation is random. We always sum across an energy range sufficient to include all the line emission. The inner circle marks $R_{\rm 500c}$, while the outer circle marks $R_{\rm 200c}$, which are $\sim 150$\,kpc and $\sim 215$\,kpc, respectively. These emission maps have a large dynamic range. The central surface brightness values reach $\sim 10^{37}$\sbunits (in white), while the majority of the halo region is much lower at $\sim 10^{34}$\sbunits, decreasing to $\sim 10^{32}$\sbunits (on average) in the outskirts, at distances comparable to the virial radius.

This galaxy has a mass similar to that of the Milky Way, with $M_\star = 10^{10.6}$\msun, a star formation rate of $\sim 7$\msunyr, and a total halo mass $M_{\rm halo} = 10^{12.0}$\msun. The supermassive black hole (SMBH) of this galaxy has transitioned into the stronger `kinetic mode', which is common at this galaxy mass in the TNG model, being the case for $\sim 2/3$ of Milky-Way mass galaxies at $z=0$ \citep{pillepich21,ramesh23}. As a result it drives a large, halo-scale outflow. In the projection shown, this outflow is visible as the bright feature towards the upper left, extending from the central galaxy across $R_{\rm 500c}$. The galaxy also has smaller scale, ongoing outflows, visible as bright inner loop-like structures on $\sim$\,tens of kpc scales. These galactic center bubbles are produced in TNG50 Milky Way-like galaxies and have features qualitatively similar to Fermi/eROSITA bubbles observed in our own Milky Way \citep{pillepich21}.

Due to the ionization state, density, and temperature in these SMBH-driven outflows, they light up as intrinsically bright \oviir features. This suggests that they may be directly observable signposts of galactic feedback. However, their strong surface brightness contrast with respect to the background CGM is partially washed out after taking into account scattering. The smoothing out of shock fronts and bubble-like structures will hinder their detection and characterization, in contrast to non-resonant lines including \ion{O}{VII}{\small (f)} and \ion{O}{VII}{\small (i)}. Observing the complete triplet is clearly advantageous, and by design any LEM pointing which captures \oviir will simultaneously cover the other two lines. This is apparent in the ratio image (right panel), which highlights the outflow-driven bubbles as regions where the surface brightness stays relatively constant or even decreases (white to light blue colors). As a result, energetic outflows light up the CGM as a whole via resonant scattering.

\begin{figure}
\centering
\includegraphics[angle=0,width=0.46\textwidth]{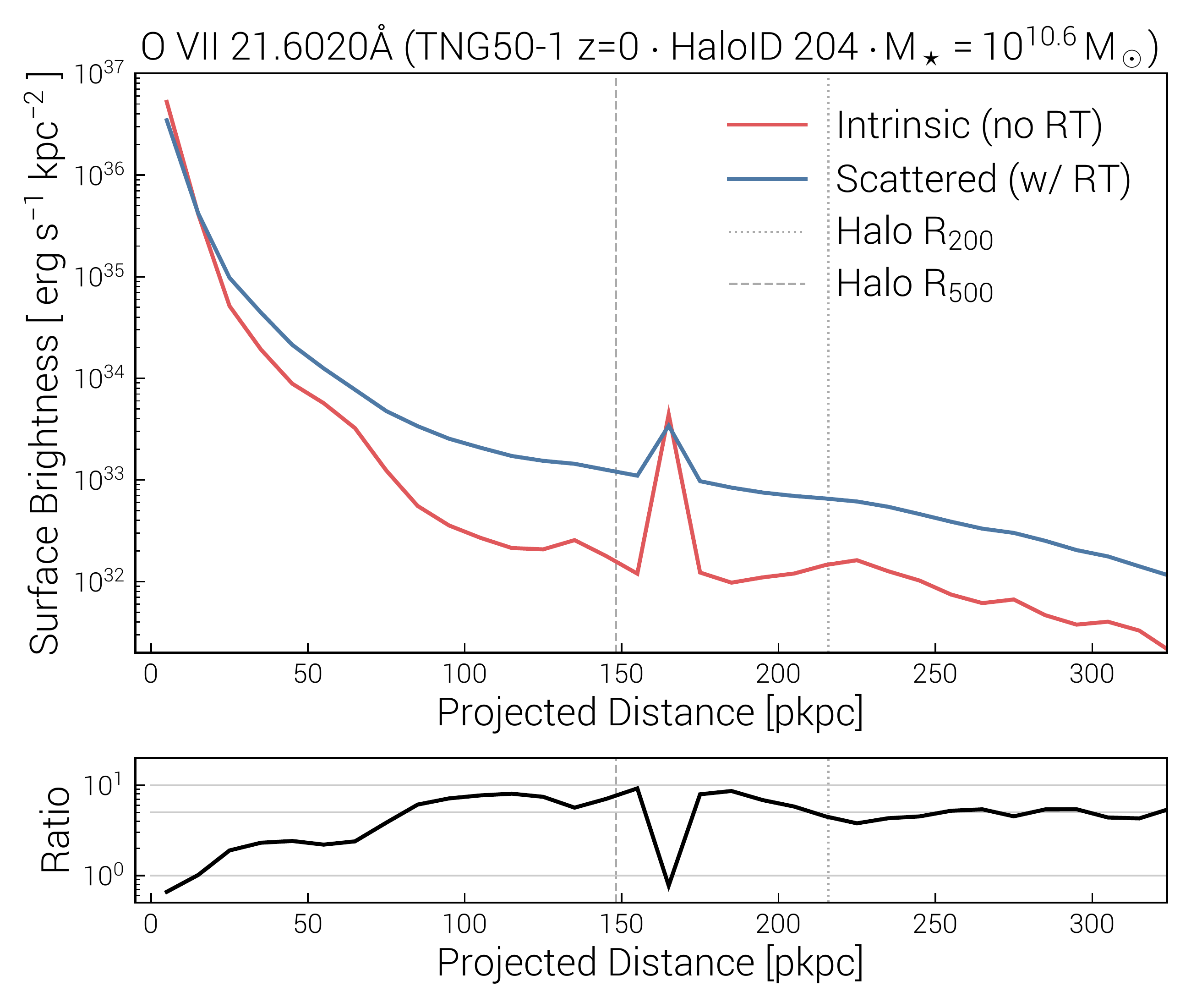}
\caption{Radial profiles of \oviir surface brightness for the same $M_\star \sim 10^{10.6}$\msun TNG50 galaxy from \ref{fig_single_maps}. The top panel contrasts the intrinsic emission, neglecting radiative transfer effects i.e. resonant scattering (red line), versus the scattered and thus observable emission (blue line). The bottom panel shows the ratio of these two profiles, indicating the surface brightness enhancement factor. This is on average a factor of $\sim$\,5 throughout the halo, and reaches up to a factor of 10, indicating the strong impact of resonant scattering. Each profile is computed in 2D projection as the total photon luminosity within each circular annulus, normalized by the area of the annuli. 
 \label{fig_single_profile}}
\end{figure}

Figure \ref{fig_single_profile} quantifies the radial surface brightness profile of \oviir emission for this same galaxy. The top panel shows the intrinsic profile neglecting scattering (red line), as well as the scattered i.e. observable profile after treating radiative transfer effects (blue line). The ratio of the scattered to intrinsic profiles, giving the effective surface brightness enhancement, is shown in the bottom panel (black line). This enhancement is actually a deficit at two locations: within the central galaxy itself, at zero projected distance, and within a satellite galaxy at a distance of $\sim 160$\,kpc, which produces a strong local enhancement of intrinsic \oviir emission. In both cases photons from these compact sources scatter outwards, illuminating their surroundings out to larger scales.

The most important feature is the redistribution of \oviir emission from the central galaxy into the extended CGM. As the luminosity at the very center of the halo is several orders of magnitude larger than in the outskirts, there is ample opportunity (i.e. available photons) to significantly enhance the observable surface brightness throughout the halo. For this halo, the enhancement ratio rises from unity at $\sim$\,10 kpc to a factor of five by $\sim$\,80 kpc, and reaches a factor of ten at $\sim$\,150 kpc ($R_{\rm 500c}$). Qualitatively speaking, the radial surface brightness profile is enhanced by a factor of several, all the way out to the virial radius. As discussed above, this enhancement can be spatially localized, e.g. greater outside of bright outflows, and the impact is not fully captured by the radial trend alone.

The intrinsic luminosity of the central galaxy in the \oviir line is $L_{\rm OVIIr} = 1.44 \times 10^{39}$\,erg s$^{-1}$, where we sum emission from within a circular aperture of 5 kpc, corresponding to 45" at $z=0.006$.\footnote{The total observable luminosity of the halo, integrated out to the virial radius, is $L_{\rm OVIIr} = 2.07 \times 10^{39}$\,erg s$^{-1}$, while the intrinsic value is $L_{\rm OVIIr} = 2.29 \times 10^{39}$\,erg s$^{-1}$, i.e. $\sim$10\% of the total \oviir emission of the halo is scattered beyond $R_{\rm 200c}$.} Scattering reduces the luminosity within this small aperture by 40\%, giving an observable luminosity of $L_{\rm OVIIr} = 8.62 \times 10^{38}$\,erg s$^{-1}$. As a point of comparison, at this same distance and within the same aperture, NGC 7213 (with a similar stellar mass) has an observed luminosity of $L_{\rm OVIIr} = 1.9 \times 10^{39}$\,erg s$^{-1}$ \citep[][with RGS]{starling05}. This suggests that our overall emission model is reasonably realistic, being neither too bright nor too faint. We consider further observational validations below.

\subsection{Diversity and variation across the galaxy population}

\begin{figure*}
\centering
\includegraphics[angle=0,width=0.9\textwidth]{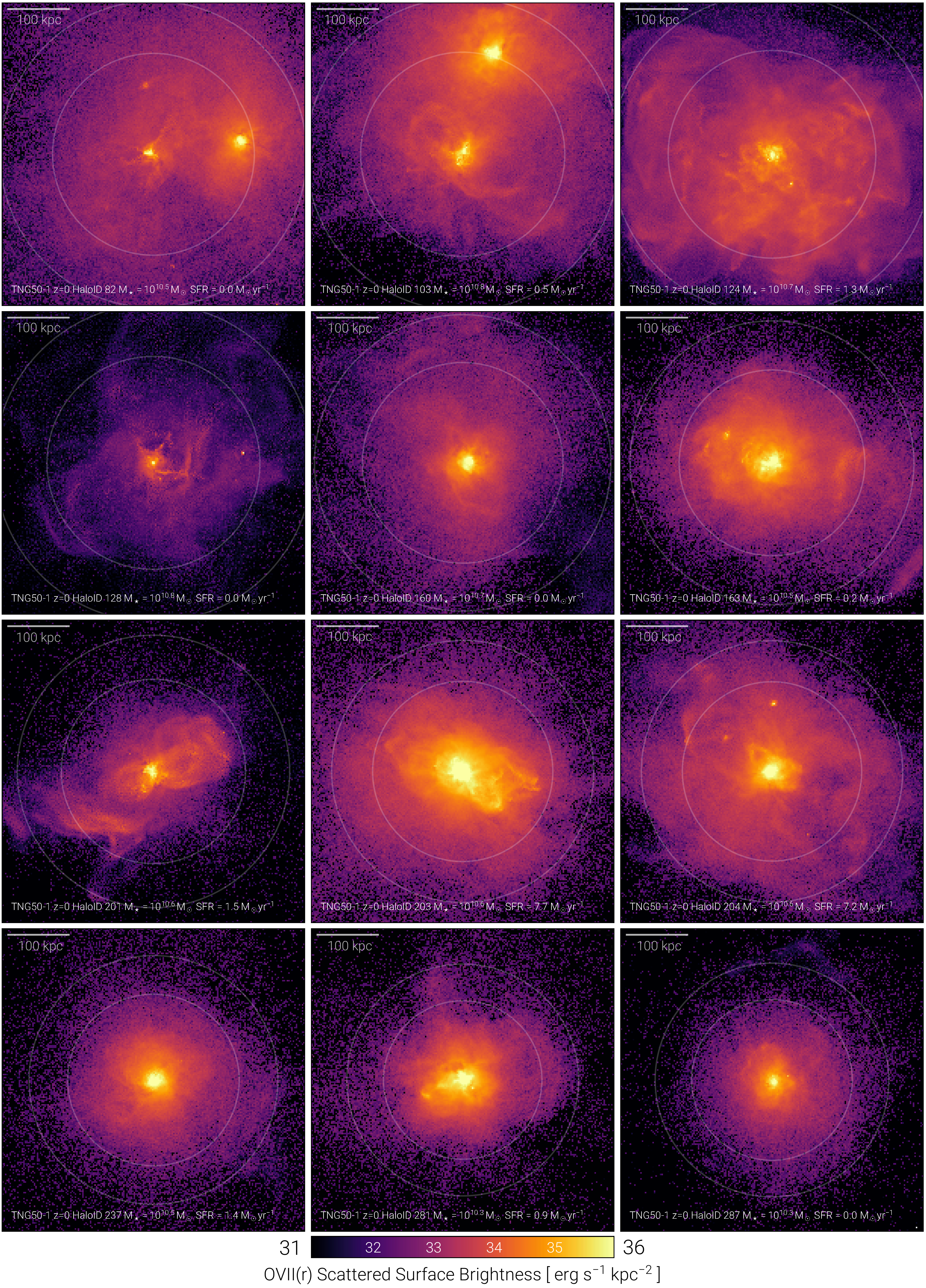}
\caption{Gallery of \oviir surface brightness maps, around twelve randomly selected TNG50 galaxies, highlighting the diversity and morphological complexity. In all cases we show the scattered, observable emission, after taking into account radiative transfer effects. Ongoing mergers, substructure, and outflows are all visible. The halo IDs, stellar masses, star formation rates, $R_{\rm 500C}$, and $R_{\rm 200c}$ scales are indicated in each case. Each image has a constant extend of 250 kpc.
 \label{fig_gallery}}
\end{figure*}

In Figure \ref{fig_gallery} we show a gallery of twelve halos, selected randomly from the full TNG50 sample. All images show the predicted observable \oviir surface brightness maps, after accounting for scattering, across a fixed field of view of 250 kpc. Stellar masses range from $10^{10.3}$\msun to $10^{10.8}$\msun, and star formation rates range from $\sim 0.1$\msunyr to $\sim 8$\msunyr. Across this subset of the sample there is already a significant diversity of structure evident. In particular, \oviir line emission from the circumgalactic medium reveals large merger events in progress (first two examples), as well as smaller and subtler substructures (last column). Many although not all of the twelve galaxies exhibit clear signatures of feedback-driven outflows. In some cases the central CGM is extraordinarily bright out to $\sim$\,10s of kpc, while other halos reach such high values only in their very cores. The surface brightness at the virial radius (outer white circle; $R_{\rm 200c}$) varies by many orders of magnitude across different halos.

\begin{figure*}
\centering
\includegraphics[angle=0,width=0.9\textwidth]{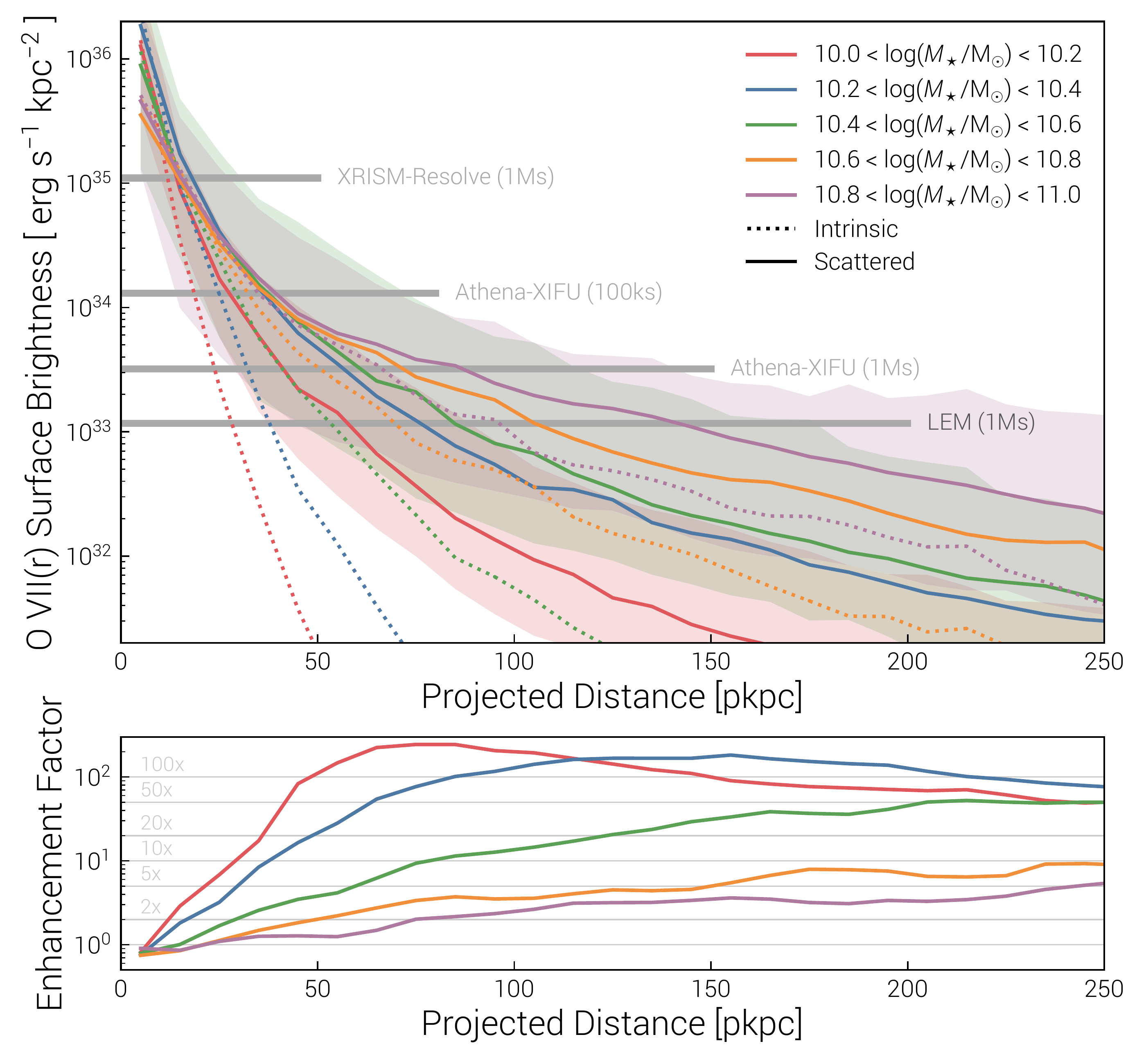}
\caption{Stacked radial surface brightness profiles of \oviir emission (top panel), and the enhancement factor due to resonant scattering (bottom panel), combining all galaxies in the TNG50 sample. The top panel compares intrinsic emission, neglecting radiative transfer i.e. scattering (dotted lines) to the scattered, and so observable, photons (solid lines). The profile of each halo is computed as before, and we then take the median of these profiles across the population, in five bins of stellar mass, from $10^{10} < M_\star / \rm{M}_\odot < 10^{11}$. Colored bands show the $16-84$ halo to halo variation for the lowest, middle, and highest mass bins. Horizontal gray lines show estimates for $5\sigma$ observational detectability levels, for the given instruments and exposure times (see text). We use the same five stellar mass bins, and colors, in the bottom panel to show the enhancement factor, i.e. the ratio of the median scattered to intrinsic profiles, as a function of projected distance. Overall, higher mass halos have brighter \oviir emitting halos. However, the scattering effect is actually more significant, in the relative sense, for lower mass galaxies. At the low mass end of our sample, $M_\star \sim 10^{10}$\msun, the peak enhancement reaches a factor of 200, on \textit{average}, at projected distances of $\sim 50-100$\,kpc. However, such low mass halos have absolute surface brightness values so low that they will be difficult to observe. Intermediate mass halos with $M_\star \sim 10^{10.5}$\msun have enhancement factors ranging from $\sim 5$ in the inner CGM to $\sim 20$ in the outer CGM. Resonantly scattered \oviir photons from the central galaxy are good news for the observability of extended CGM emission.
 \label{fig_stacked_profiles}}
\end{figure*}

Given this significant diversity across the population, we next assess the average impact of resonant scattering on observable \oviir emission. Figure \ref{fig_stacked_profiles} shows radial \oviir surface brightness profiles, contrasting intrinsic (dotted lines) versus scattered (solid lines) emission (top panel). We stack all galaxies in the sample into five bins of stellar mass, from $10.0 < \log{(M_\star / \rm{M}_\odot)} < 11.0$, where individual profiles are constructed as before, and we median combine the profiles of halos in each bin. Shaded regions, included only for three bins for visual clarity, show the $16-84$ percentile halo to halo scatter.

As a function of galaxy stellar mass, the \oviir surface brightness of the extended CGM increases for more massive galaxies. At a fixed distance of $100$\,kpc, from $10^{32}$\sbunits at $M_\star \sim 10^{10.0}$\msun, to $2 \times 10^{33}$\sbunits at $M_\star \sim 10^{10.8}$\msun. To some degree this reflects the increasing size of the gaseous halos, as we plot distances throughout in physical kpc and do not normalize by the virial radius. The central luminosities have the opposite trend -- less massive galaxies have brighter cores, on scales of $\lesssim$\,30 kpc. As a result, the radial surface brightness profiles of more massive galaxies are shallower, i.e. flatter and more extended, than their lower mass counterparts.

At all stellar masses we consider, scattering boosts the \oviir surface brightness profiles of the extended CGM.\footnote{We have explicitly checked the importance of including the velocity field of the gas. To do so, we assume that both the emitting and scattering media are at rest. In this case, the scattered emission is substantially brighter, from $\sim$\,10\% to a factor of two, depending on mass and distance (not shown). As thermal broadening is not a significant factor for \oviir, velocities are important for shifting photons out of resonance. With the velocity field zeroed, there is significantly more scattering. This demonstrates the importance of including the gas kinematics in order to not overestimate the impact of scattering.} With horizontal gray lines, we include estimates for the surface brightness values reachable for upcoming X-ray instruments: XRISM-Resolve, Athena-XIFU, and LEM. For the first two cases, we adopt a $100\,\rm{km\,s^{-1}}$ line width, a $5\sigma$ detection significance, include \textsc{xspec wabs} Galactic absorption with $N_{\rm H} = 1.8 \times 10^{20} \rm{cm}^{-2}$, and assume a single field of view pointing with the given exposure time is binned \citep[adopted from][]{wijers22}.\footnote{Athena X-IFU values derived before the Athena reformulation as of 2022.} For the case of LEM we estimate the total background rate within a $4$\,eV band at the location of the redshifted \oviir line to be $4.93 \times 10^{-5}$\,cnt s$^{-1}$ arcmin$^{-2}$, including three thermal foreground components, the cosmic X-ray background without removing any bright sources, and the conservative estimate of the LEM instrumental background. We then estimate the LEM 5$\sigma$ line sensitivity assuming an effective area of $1700$\,cm$^2$ \citep{kraft22}, for a source filling the LEM field of view. As this threshold reaches $\lesssim 10^{33}$\sbunits, the distance out to which each profile can be detected becomes much greater after accounting for scattering. We also emphasize the mapping capabilities of LEM: given that the FoV is $\sim 40$ times larger than Athena X-IFU, the latter would require a mosaic of roughly this number of pointings to cover a field-filling CGM to the same depth, requiring substantially more exposure time than indicated in the Figure.

The lower panel of Figure \ref{fig_stacked_profiles} quantifies this enhancement, as before, taking the ratio of the scattered to intrinsic radial profiles. Horizontal gray lines mark enhancements of 2x, 5x, 10x, 20x, 50x, and 100x, as indicated. Resonant scattering leads to a larger enhancement for extended surface brightness levels in lower mass galaxies. Remarkably, the \textit{average} enhancement reaches a peak factor of 200, at $\sim 50-100$\,kpc scales around $M_\star \sim 10^{10.0-10.2}$\msun galaxies (red lines). However, the observable surface brightness is ultimately much lower for lower mass galaxies, i.e. although the relative scattering-induced enhancement is impressive, it is largely irrelevant from an observational detectability point of view. The most massive galaxies, up to $M_\star = 10^{11}$\msun, are still the brightest beyond $\gtrsim 20-30$\,kpc, as expected given the increasing density of the CGM as well as the relatively flat abundance of \ion{O}{VII} across this halo mass range -- \ion{O}{VIII} starts to dominate just beyond $M_\star \sim 10^{11}$\msun\citep{nelson18b}.

Particularly in the two lower mass bins, we can see that the enhancement factor is not monotonic with distance, but instead has a maximum at intermediate scales. Namely, the average enhancement within $\lesssim 10$\,kpc is less than unity, and begins to increase monotonically with increasing distance. For galaxies with $M_\star \sim 10^{10.0-10.2}$\msun, a peak occurs at $\sim$\,70 kpc, which is $\sim 0.5 R_{\rm 200c}$ of these halos. For galaxies with $M_\star \sim 10^{10.2-10.4}$\msun, the peak is broad and occurs at $\sim$\,150 kpc, which is $\sim 0.8 R_{\rm 200c}$, while for the three most massive bins the enhancement factor monotonically rises within the entire 250 kpc range that we consider here.

\begin{figure*}
\centering
\includegraphics[angle=0,width=0.7\textwidth]{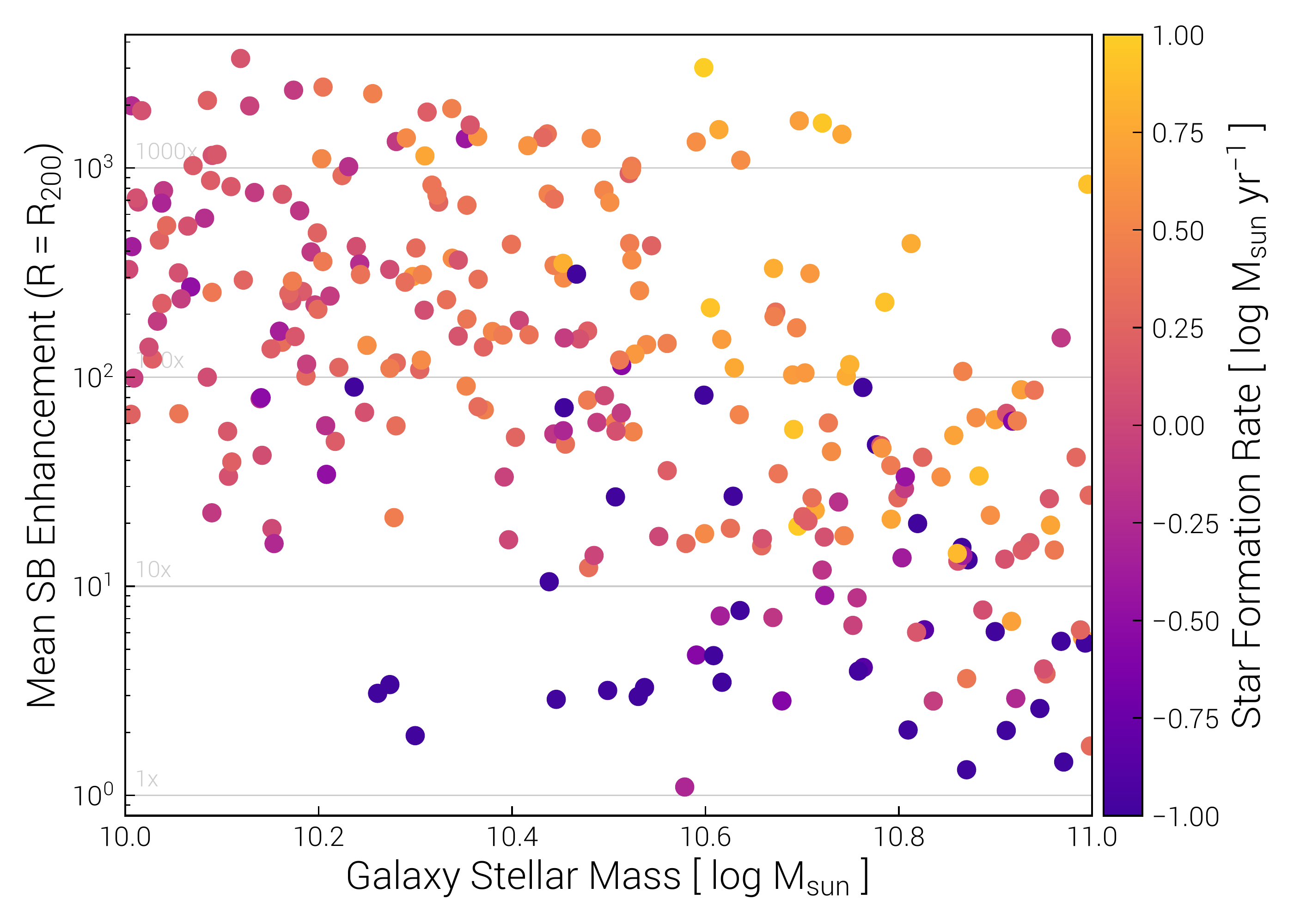}
\includegraphics[angle=0,width=0.32\textwidth]{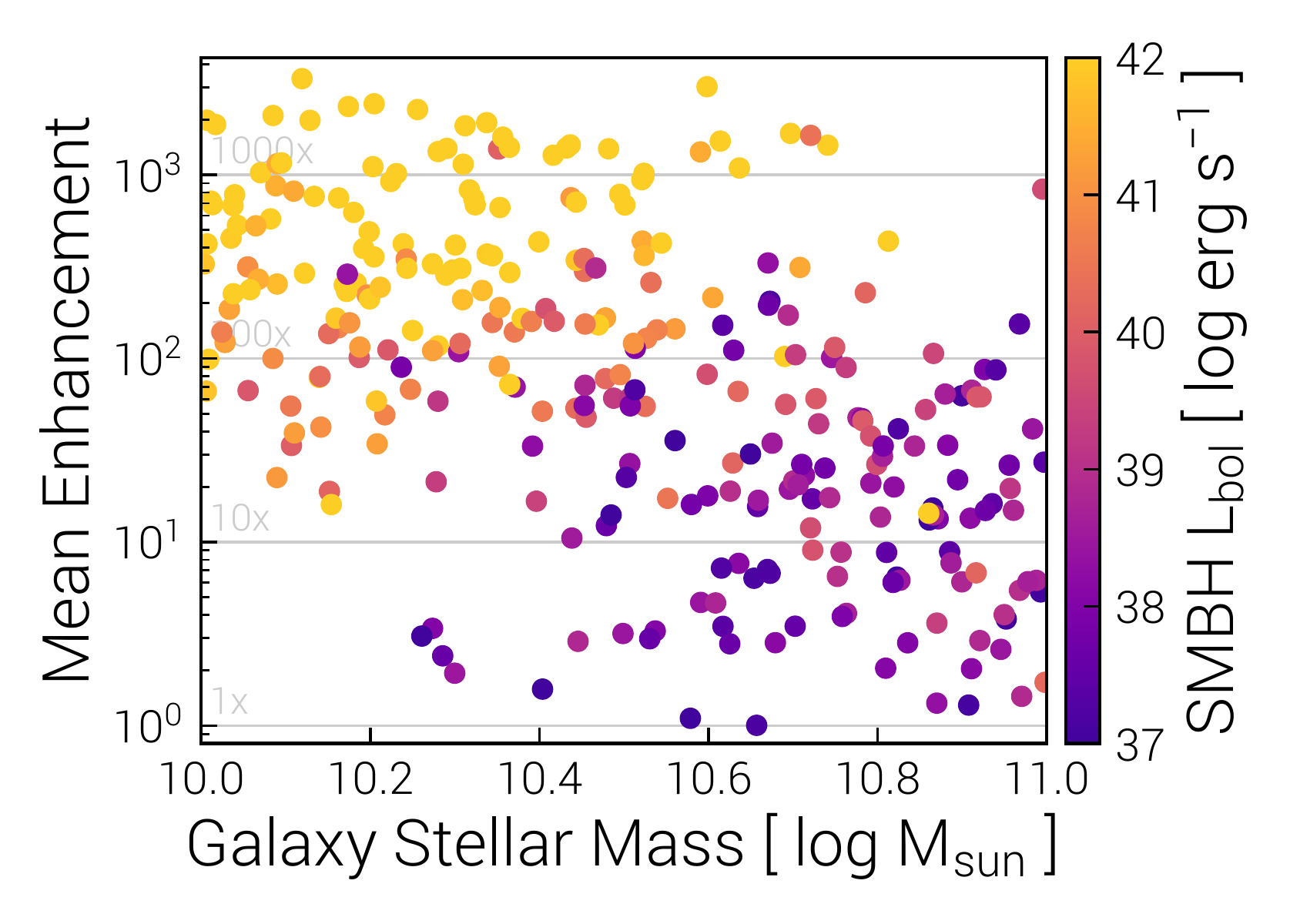}
\includegraphics[angle=0,width=0.32\textwidth]{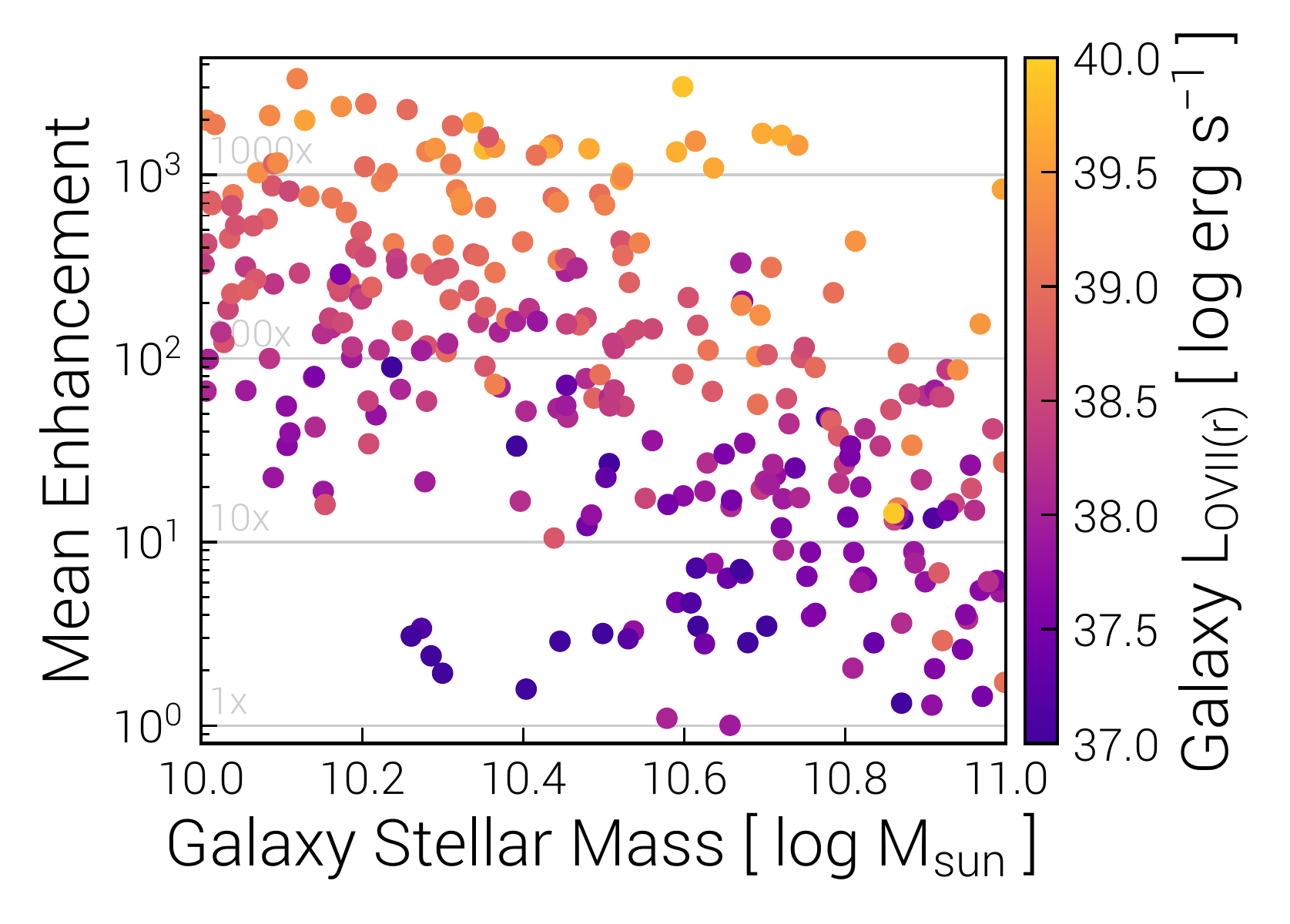}
\includegraphics[angle=0,width=0.32\textwidth]{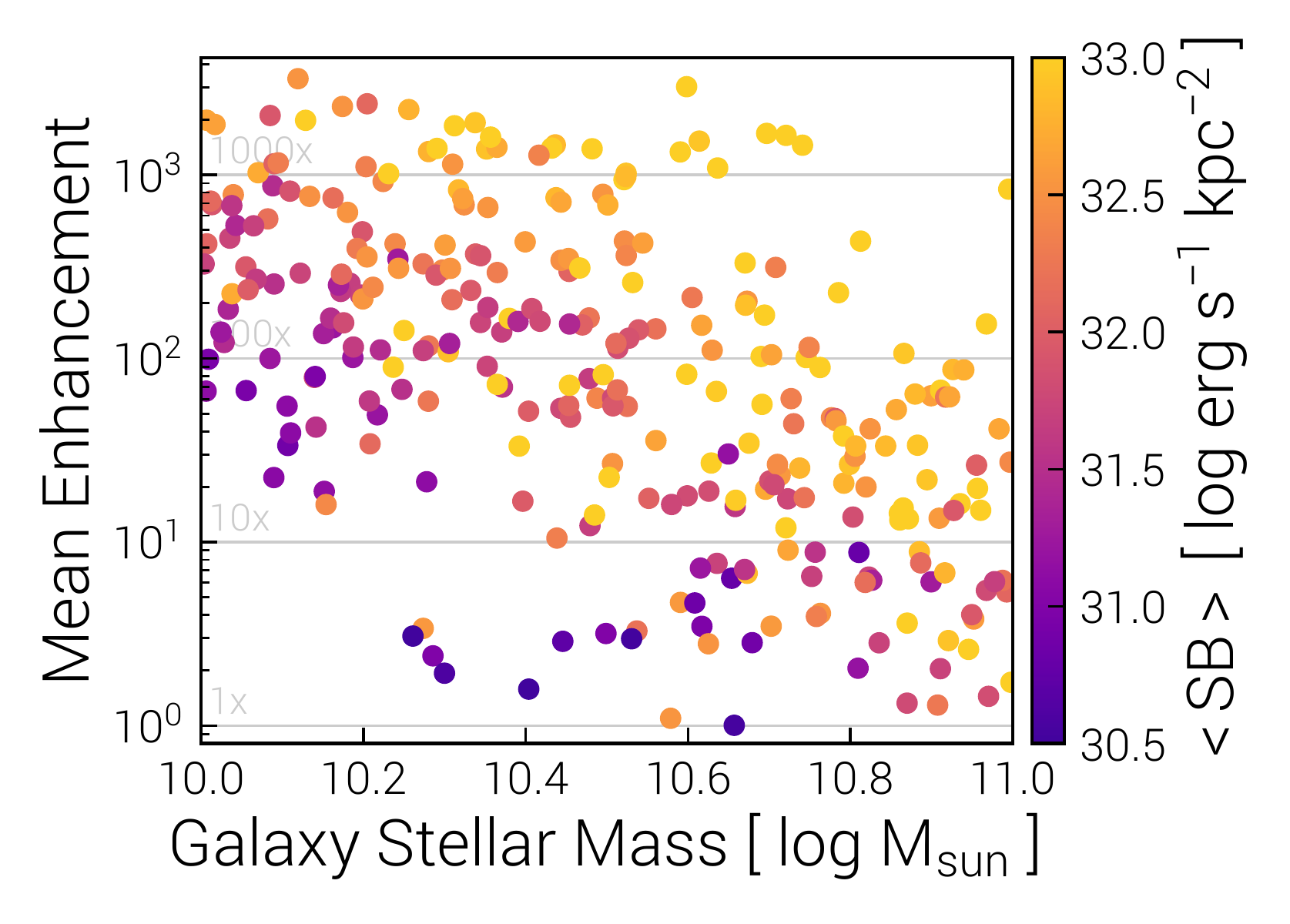}
\caption{The average \oviir surface brightness enhancement due to resonant scattering at the virial radius, as a function of galaxy stellar mass, for the entire TNG50 sample. For each halo we take the ratio of the scattered to intrinsic maps, and then compute the mean ratio within $0.95 < R/R_{\rm 200c} < 1.05$. Individual halos are shown as circular markers, colored according to galaxy star formation rate (main, top panel), SMBH bolometric luminosity (lower left panel), central galaxy \oviir luminosity (lower middle panel), and mean, scattered \oviir surface brightness at $R_{\rm 200c}$ (lower right panel). On average, the enhancement factor due to resonant scattering decreases with increasing galaxy mass, from $\sim$\,hundreds to $\sim$\,a few across our mass range. However, the scatter at fixed mass is comparable to the overall mass trend. At fixed stellar mass, galaxies with higher SFRs, AGN luminosities, and central $L_{\rm OVIIr}$ have larger surface brightness enhancements, which correspond to larger observable surface brightness values.
 \label{fig_enhancement_vs_mass}}
\end{figure*}

The most important quantitative result of Figure \ref{fig_stacked_profiles} is the enhancement factor on CGM scales for intermediate $M_\star \sim 10^{10.5}$\msun mass galaxies, where \oviir emission would potentially reach observable levels. For galaxies with this stellar mass (green lines), the enhancement rises from a factor of 2 at $\sim$\,30 kpc, to a factor of 5 at $\sim$\,60 kpc, to a factor of 10 at $\sim$\,100 kpc. It eventually reaches a factor of 50 at even larger distances of $\sim$\,200 kpc. The relevance of this depends on the observability of an absolute surface brightness level of $\sim 10^{32}$\sbunits in \oviir line emission. For more massive galaxies with stellar mass $M_\star \sim 10^{10.7}$\msun (orange lines), the enhancements due to scattering are less impressive, although the actual observable surface brightness values are higher. For example, on average galaxies have an enhancement factor of 5 at $\sim$\,100 kpc, where the predicted surface brightness is then $\sim 10^{33}$\sbunits. Our most massive bin with $M_\star \sim 10^{10.9}$\msun (purple lines) has essentially the same predicted median radial \oviir surface brightness profile, as the increasing halo mass and thus intrinsic emission is counterbalanced by a decreasing impact of resonant scattering.

Overall, these results quantify our population level prediction that scattering of the \oviir line significantly boosts the observable surface brightness values, as a function of galaxy stellar mass and projected distance. What remains unclear is the level of halo to halo variation, and whether other observables of a galaxy may indicate, a priori, that the \oviir emitting halo of that system will be brighter.

\subsection{Connection to galaxy properties}

Figure \ref{fig_enhancement_vs_mass} shows the mean \oviir surface brightness enhancement as a function of galaxy stellar mass, at a particular choice of distance: the virial radius $R_{\rm 200c}$. In each of the four panels, all galaxies in the sample are shown with individual colored symbols, where color corresponds to galactic star formation rate (main panel), bolometric luminosity of the central SMBH/AGN (lower left panel), central \oviir luminosity of the galaxy itself, within a 10 kpc aperture (lower middle panel), and mean, scattered \oviir surface brightness at $R_{\rm 200c}$ (lower right panel).\footnote{These statistics, particularly median values, depend to some degree on pixel size (PSF), and we adopt an angular resolution of 60" at $z=0.01$, corresponding to a smoothing scale of $\simeq 12.8$\,kpc. In order to minimize shot noise from our Monte Carlo radiative transfer, we select this value to be four times worse than the LEM concept, which is optimized for characterizing the CGM due to its large grasp and high spectral resolution \citep{kraft22}.}

The overall trend with mass is monotonically decreasing, as previously seen: the impact of scattering is largest for the lower mass galaxies. However, the scatter is enormous. At intermediate masses, halo to halo variation in the surface brightness enhancement factor spans the entire range, from essentially unity to a factor of 1000. In the top panel, color indicates log of galaxy star formation rate, and a clear correlation exists. At fixed stellar mass, galaxies with higher star formation rate (SFR) in their centers have larger enhancements of \oviir surface brightness at their virial radii. This effect is strong: systems with $\rm{SFR} \lesssim 0.2$\msunyr have hardly any enhancement from scattering, while systems with $\rm{SFR} \gtrsim 5$\msunyr have the largest.

The three lower panels of Figure \ref{fig_enhancement_vs_mass} show that the \oviir surface brightness enhancement is also connected to other galactic properties. Namely, to the luminosity of the central AGN (lower left panel), which is computed assuming an accretion rate dependent radiative efficiency \citep[see][]{nelson19a,churazov05}
\begin{equation}
L_{\rm bol} \, = \,
\left\{\!\begin{aligned}
  \, &\epsilon_r\, \dot{M}_{\rm SMBH} \, c^2 \quad &; \quad \lambda_{\rm edd} \ge 0.1\\[1ex]
  \, &10\, \lambda_{\rm edd}^2\, L_{\rm Edd} \quad &; \quad \lambda_{\rm edd} < 0.1
\end{aligned}\right\} ,
\end{equation}

\noindent where $\lambda_{\rm edd}$ is the Eddington ratio, and no obscuration is considered. Higher $L_{\rm bol}$ therefore indicate increased SMBH accretion rates, tracing more massive/denser gas reservoirs in the galaxy, which also then lead to higher SFRs. That is, the correlation with AGN luminosity does not necessarily imply a physical causation. We also note that although the TNG model does treat -- albeit in a simplified manner -- radiation from AGN, particularly its impact on gas cooling physics, we do not include AGN radiation in our radiative transfer simulations. Regardless, at fixed stellar mass, galaxies with higher SFRs and/or AGN luminosities (as well as higher SMBH masses; not shown) have the greatest scattering-induced enhancements of \oviir surface brightness.

As expected, the enhancement also correlates directly with $L_{\rm OVIIr}$ of the central galaxy itself (Figure \ref{fig_enhancement_vs_mass}, lower middle panel). The scaling is almost linear: if the central galaxy produces ten times more \oviir photons, then the resulting surface brightness at the virial radius is of order ten times higher. We also evaluate the distribution of line-center optical depth for this line (not shown), finding a large variation across halos with similar mass. While one halo can have $\tau \sim 100$ within the central galaxy, other cases can have $\tau \sim 1$, indicating that significant differences in the abundance of \ion{O}{VII} exist at the population level, and partially drive the scatter seen here.

On the other hand, there is no correlation of surface brightness enhancement with the parent dark matter halo mass $M_{\rm 200c}$ at fixed $M_\star$ (not shown). There is also no strong correlation with SMBH mass at fixed stellar mass (not shown). For the lower half of our mass range, $10^{10.0} < M_\star / \rm{M}_\odot < 10^{10.5}$, stellar feedback dominates in the TNG simulations. In this regime, higher mass halos have more total gas and higher densities, leading to larger optical depths as well as overall emission. For the upper half of our mass range, $10^{10.5} < M_\star / \rm{M}_\odot < 10^{11.0}$, where SMBH feedback begins to dominate in TNG, this is not necessarily the case, as more massive halos host more massive black holes which have injected larger cumulative amounts of kinetic energy, lowering halo-scale gas fractions \citep{zinger20,davies20,ayromlou23}. The lack of correlation between surface brightness enhancement at the virial radius and halo mass suggests that these effects are subdominant with respect to the total core luminosity, i.e. in the halo center.

Finally, the lower right panel of Figure \ref{fig_enhancement_vs_mass} demonstrates an important point. It colors each symbol by the mean, scattered surface brightness of \oviir emission at the virial radius. A clear trend with enhancement factor is present. This need not necessarily be the case, for instance if the intrinsically dimmest halos were enhanced the most due to scattering, leaving their observable surface brightness values still too low to be observed. However, we see that larger enhancement factors imply larger observable surface brightness values. As a result, the previous correlations with galaxy SFR, AGN luminosity, and central \oviir luminosity are all promising tracers of galaxies whose extended \oviir emission will be brightest.

\subsection{Dependence on distance and location within the CGM}

\begin{figure}
\centering
\includegraphics[angle=0,width=0.48\textwidth]{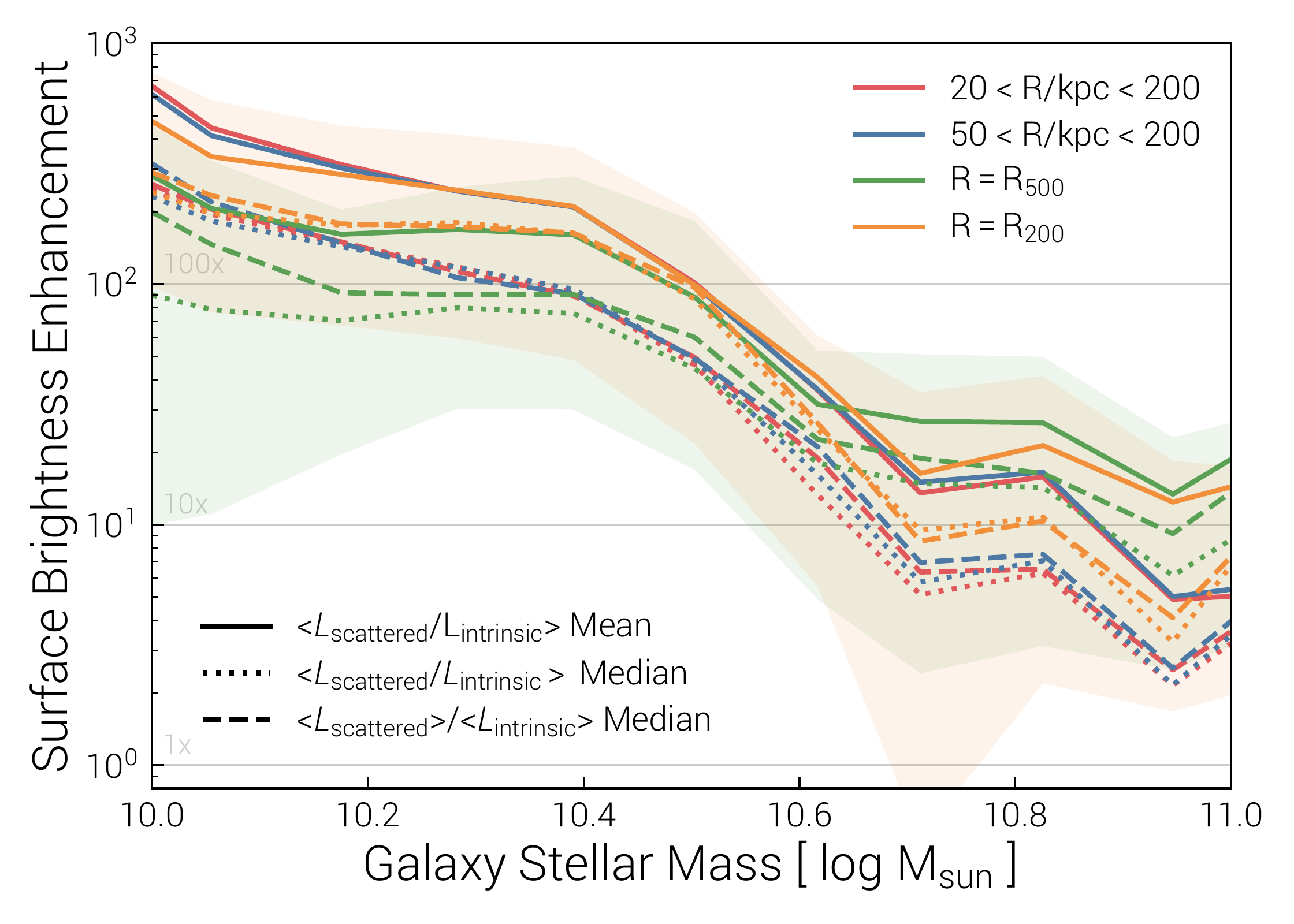}
\caption{The surface brightness enhancement factor due to resonant scattering, as a function of galaxy stellar mass. We compare four radial ranges spanning the circumgalactic medium (different line colors). These are: the bulk of the halo excluding the center (red), the outer CGM only (blue), exactly at $R_{\rm 500c}$ (green), and exactly at $R_{\rm 200c}$ (orange, as previously in Figure \ref{fig_enhancement_vs_mass}). We also compare three different statistics applied to the surface brightness maps: the mean ratio of scattered to intrinsic pixels in the given radial range (solid), the median ratio of the same (dotted), and the ratio of the median scattered pixel value to the median intrinsic pixel value. Differences as a function of distance are minimal: all four radial ranges have similar enhancements, although the outer halo has the least enhancement at low stellar mass, and the most enhancement at high stellar mass. The mean enhancement at the map level is higher than the median, suggesting that localized regions are preferentially boosted.
 \label{fig_enhancement_vs_rad}}
\end{figure}

The degree to which scattering boosts \oviir surface brightness varies throughout the halo as a function of distance. Figure \ref{fig_enhancement_vs_rad} again shows the \oviir surface brightness enhancement as a function of galaxy stellar mass, where we now visualize the result with the running median trend, instead of individual markers. This allows us to compare four different distance regimes: the entire circumgalactic medium excluding the central 20 kpc (red), the outer CGM only ($50\,\rm{kpc} < R < 200\,\rm{kpc}$; blue), at $R_{\rm 500c}$ of the halo (green), and at $R_{\rm 200c}$ of the halo (orange, as shown previously). The median trends of \oviir surface brightness enhancement are relatively insensitive to this choice. In all cases, the median across the galaxies decreases rapidly with mass, from $\sim\,200 - 600$ at $M_\star = 10^{10.0}$\msun to $\sim\,3 - 20$ at $M_\star = 10^{11.0}$\msun. Across the four radial ranges considered, variation in the median is at the factor of two level at most. At low stellar masses, the halo outskirts have the smaller enhancements in comparison to the inner halo. In contrast, at high stellar masses, the halo outskirts have the larger enhancements.

In Figure \ref{fig_enhancement_vs_rad} we also compare the enhancement computed in three different ways. First, as the mean of the ratio of the scattered luminosity to intrinsic luminosity, across all pixels in the relevant region of the surface brightness maps (solid lines). Second, as the median of the ratio of the same regions, i.e. pixel by pixel (dotted lines). Third, first taking the median surface brightness for scattered pixels, and likewise for intrinsic pixels, in the relevant region, and then their ratio. Overall, the enhancement factor does not strongly depend on this choice.\footnote{The first approach is consistent with Figure \ref{fig_enhancement_vs_mass}, while the last approach is more similar to the ratio of the radial surface brightness profiles as in Figure \ref{fig_stacked_profiles}. We also note that taking the ratio of the mean scattered pixel value to the mean intrinsic pixel value, which is the same as taking the ratio of the total scattered luminosity to total intrinsic luminosity in the given radial range, is exactly consistent with Figure \ref{fig_stacked_profiles}. In this case, the surface brightness enhancement is consistent with all the other measurements at large radii, while radial ranges including the inner halo have much lower enhancement factors, being dominated in luminosity by the bright central regions.} The mean map enhancement is generally larger than the median, suggesting that localized regions within the CGM are preferentially boosted by scattering, even at a given, fixed distance such as $R_{\rm 500c}$. However, discreteness effects and shot noise in our Monte Carlo radiative transfer technique will artificially increase this effect to some degree, especially at smaller pixel sizes. The median values represent sky area weighted values, and median surface brightness enhancements tend to be a factor of $\sim$\,2 lower than mean values, regardless of stellar mass. At $M_\star \sim 10^{10.5}$\msun the median enhancement factors are up to $50$\% lower than in the mean.

\subsection{Observational checks and modeling sensitivities}

We have seen that the impact of scattering depends strongly on the available luminosity in the bright, central region of the halo. This could either arise from the increasing dense gas of the CGM towards the halo center, with temperatures resulting from structure formation and of order the virial temperature \citep{nelson16,ramesh23}. Alternatively, it could arise from the hot, star-forming interstellar medium of the galaxy, or emission from the AGN, itself. In our case, the former is a direct output of the simulation coupled to our \textsc{CLOUDY} based emission model, and has no flexibility, while the latter is an ingredient designed to be controlled via the boost $b$-parameter. We explore these aspects here, starting with a comparison with available observational constraints.

\begin{figure}
\centering
\includegraphics[angle=0,width=0.48\textwidth]{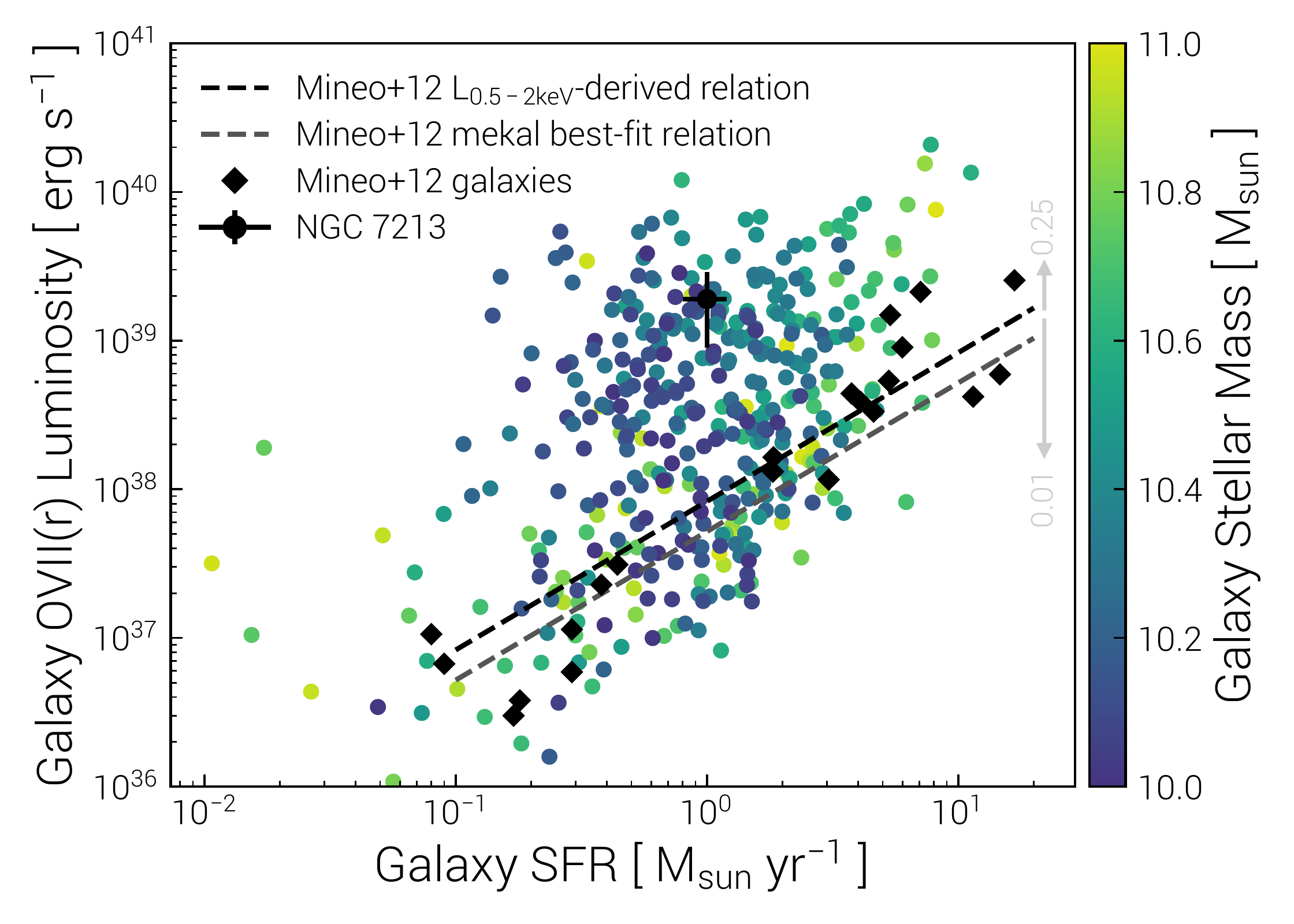}
\caption{Scaling relation between central galaxy \oviir luminosity and star formation rate. Our predictions for the full TNG50 sample given our fiducial emission model are shown as colored circles, color denoting stellar mass. We measure simulated luminosities within roughly the effective radius (here, the stellar half mass radius). We include a single direct observational measurement, for NGC 7213 (see text). In addition, we include the observational results of a 21 galaxy sample from \citet{mineo12}, showing both the individual measurements (black diamonds) and best-fit relations (dashed lines). This data measures the entire $0.5 - 2$\,keV soft-band X-ray luminosity, and we rescale these values downwards by a factor of ten as a rough estimate of the fractional contribution from the \oviir line alone. As a result, and due also to the significant uncertainties involved in the observational inferences, the \citet{mineo12} comparison should be understood as a qualitative, order of magnitude assessment, rather than a quantitative apples-to-apples test (see text).
 \label{fig_obs}}
\end{figure}

Figure \ref{fig_obs} shows the scaling relation between galaxy \oviir luminosity and star formation rate. The TNG50 predictions, given our emission modeling, are shown as circles for the full sample, with color indicating stellar mass. We overplot the observational sample of \citet{mineo12} which includes 21 nearby, late-type galaxies with SFRs ranging from $0.1 - 20$\msunyr and stellar masses $3 \times 10^8 - 6 \times 10^{10}$\msun. We show the individual measurements (black symbols), as well as two best-fit relations (dashed lines).

A critical caveat exists: \citet{mineo12} measure the entire $0.5 - 2$\,keV soft X-ray luminosity, and not the \oviir line luminosity alone. However, emission from the \oviir line is a fractional contribution to the broadband $L_{\rm 0.5-2 keV}$ total. Knowing this fraction we could convert the observed broadband values into \oviir line values, and vice versa. We have therefore used \textsc{Cloudy} to compute this fraction for a single-zone case, for solar as well as one third solar metallicity gas, where the results are similar. The (\oviir / $0.5-2$ keV) flux ratio has a strong temperature and density dependence, peaking at $\sim 0.25$ for temperatures $T_{\rm gas} \sim 10^{6.0 - 6.2}$\,K and densities $n > 10^{-3}$\,cm$^{-3}$. That is, the \oviir line can be a large component of the soft broadband emission.\footnote{This would imply that even broadband $0.5-2$\,keV luminosities predicted from theoretical models, if spatially localized e.g. to a galaxy or a particular regime of the CGM, could be substantially modified by scattering effects. Similarly, comparison of radial $0.5-2$\,keV profiles \citep[e.g. as recently undertaken with eROSITA/eFEDS;][]{comparat22,chadayammuri22} must account for scattering of the individual lines within this band, to avoid the erroneous interpretation that simulated profiles are too steep.} Outside of this range, for $T < 10^{5.9}$\,K or $T > 10^{6.4}$\,K or $n \leq 10^{-4}$\,cm$^{-3}$, the fractional contribution drops rapidly to negligible amounts. However, the hot interstellar medium and central dense CGM gas span a wide range of physical densities and temperatures. We have also preformed a calculation with the APEC collisional ionization model, considering all non-starforming gas within $<0.1 R_{\rm 200c}$ within TNG50-1 galaxies across our mass range. In this case, we find that the \oviir line flux is $\sim$\,10-15\% of the $0.5 - 2$\,keV total. However, this case excludes the hot ISM, and it is not clear if this is a lower limit, or how robust this fraction is.

We therefore proceed by adopting a reasonable fractional value of 0.1, and reduce the \citet{mineo12} broadband luminosities by this factor for a more direct comparison against the simulations in Figure \ref{fig_obs}. We simultaneously show, with gray arrows, the vertical shift had we instead assumed either 25 percent or 1 percent. We caution, however, that the resulting comparison is intended in the qualitative sense, i.e. as an order of magnitude assessment, rather than as a robust quantitative assessment. In addition to the uncertain conversion from line to broadband luminosity, the observational analysis of \citet{mineo12} also involves several additional steps. To isolate the emission of the hot ISM, they must remove many other sources of emission, including compact X-ray sources and counts from extremely bright compact X-ray sources, including PSF effects, as well as instrumental and cosmic X-ray backgrounds, unresolved high-mass and low-mass X-ray binaries, and other types of active/young stars. In addition, we measure luminosities within an aperture of the stellar half mass radius, which may not directly coincide with the aperture used in the observational analysis. Given the analysis complexities, differing samples, and different physical measurements, we do not suggest any direct nor quantitative comparison. Instead, we treat the \citet{mineo12} relations between $L_{\rm 0.5-2 keV}$ and SFR as a qualitative guide. 

Despite the caveats, the comparison is useful. First, the observations of \citet{mineo12} show a positive correlation between soft X-ray luminosity of the halo gas and galaxy SFR, which is also present in our model \citep[see also][]{truong20}. However, this face-value comparison also suggests that, if anything, TNG50 central \oviir luminosities may be larger than in reality. This is an important caveat to our results on the impact of resonant scattering. If the simulated central luminosities in the \oviir line are too large (small), the relative impact of scattering will be over (under) estimated. Future data providing large statistical measurements of galaxy \oviir luminosity as a function of mass, including from LEM itself, are needed to help resolve this source of uncertainty.

\begin{figure}
\centering
\includegraphics[angle=0,width=0.46\textwidth]{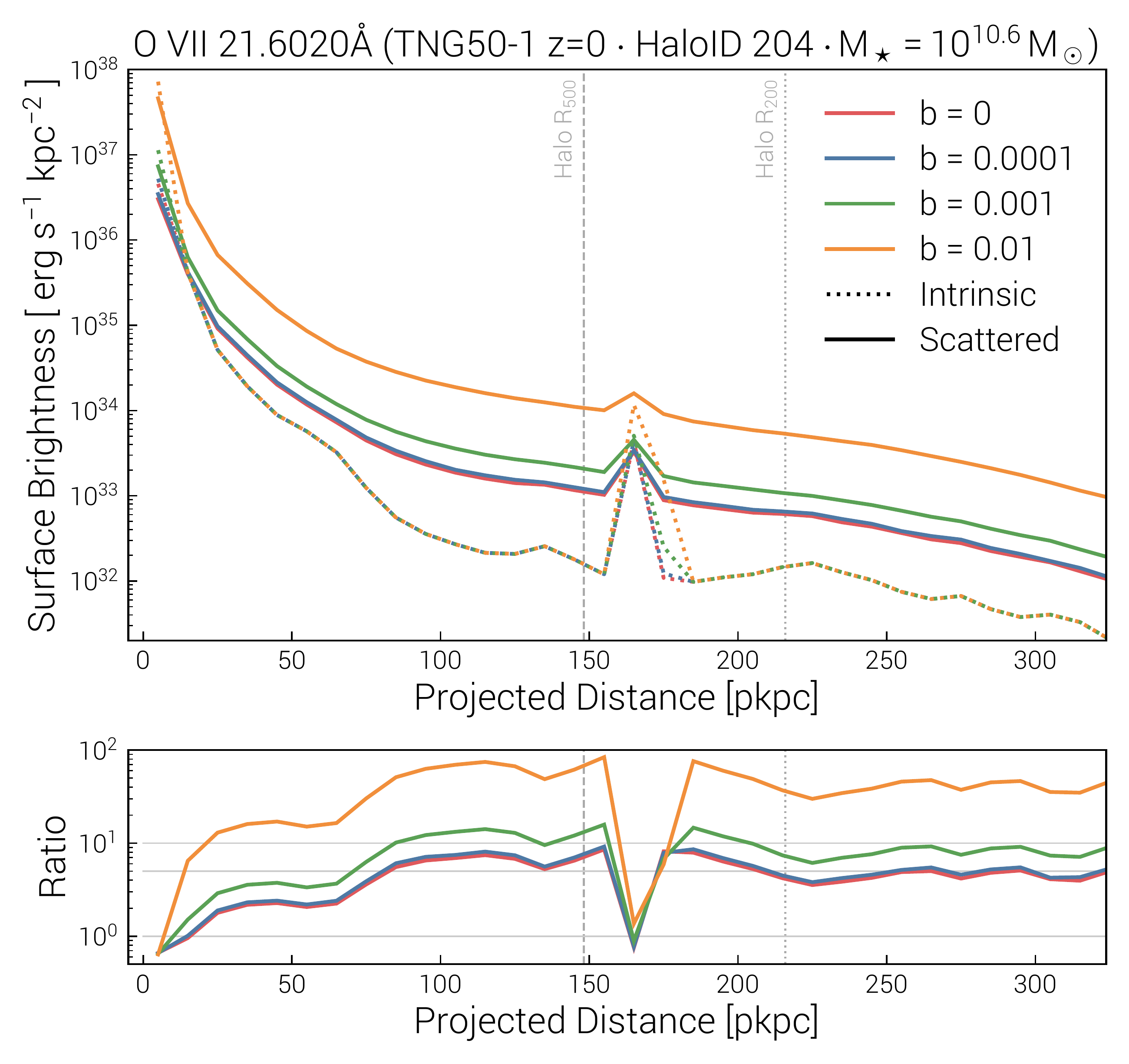}
\caption{Impact of the \oviir emission component arising from star-forming gas in the galaxy itself, i.e. from the hot interstellar medium and/or a central AGN. Here we show results for a single TNG50 galaxy, the same as in Figure \ref{fig_single_maps}. We compare the projected radial surface brightness profile for four different values of $b$, our free parameter which scales the strength of this component. For $b=0$ (red line), star-forming gas does not emit. For our fiducial choice of $b=10^{-4}$ (blue line, as in all previous analysis), the additional contribution is vanishingly small. For $b=10^{-3}$ (green line) and especially $b=10^{-2}$ (orange line) the increase of the intrinsic \oviir emission (dotted lines) is sufficient to substantially enhance the observable surface brightness (solid lines) across the entire halo. By design, this $b>0$ emission is confined to the central galaxy at $\lesssim 10$\,kpc together with the satellite system located at $\sim 160$\,kpc. At our adopted fiducial value of $b$, this component has no significant importance for our emission model.
 \label{fig_bparam}}
\end{figure}

Emission from diffuse, circumgalactic gas in \oviir has not been observationally detected. Measurements of \oviir luminosity from galaxies themselves are, indeed, limited, which is why we focused on the broadband \citet{mineo12} results above. In terms of \oviir emission, we have previously discussed NGC 7213, which has an observed luminosity of $L_{\rm OVIIr} = 1.5 \times 10^{39}$\,erg s$^{-1}$ \citep[][with RGS]{starling05,salvestrini20}. This is a S0 at $D=23$\,Mpc, with a bolometric $L_{\rm AGN} = 1.7 \times 10^{43}$\,erg s$^{-1}$, a SMBH with mass $\sim 10^8$\msun at $\lambda_{\rm edd} \sim 10^{-3}$, a SFR of $\sim 1.0$\msunyr, and a strong outflow. A significant fraction of the \oviir emission from this galaxy may be due to the AGN itself, and/or to an AGN-driven outflow. We include NGC 7213 in Figure \ref{fig_obs}, as a potentially `normal' galaxy against which we can compare. In contrast, NGC 253 is a nearby SAB starburst at $D=4$\,Mpc which drives a hot superwind, similar to M82. Summing up all flux across four available spatial regions, it has a low luminosity of \mbox{$L_{\rm OVIIr} = 9.0 \times 10^{36}$\,erg s$^{-1}$} \citep{bauer07}. Finally, \citet{liu12} present a sample of nine nearby star-forming galaxies, with luminosities ranging from $L_{\rm OVIIr} = 1 - 26 \times 10^{37}$\,erg s$^{-1}$. Ultimately, the realism of our \oviir emission model, and thus our results on the important impact of resonant scattering, require future X-ray data to fully assess.

A more accessible sanity check with data is the comparison of total broadband X-ray luminosity, i.e. in the $0.5-2$\,keV energy range. Our line emission modeling in this work does not enable such a direct comparison, however we note that the TNG model outcome, and the TNG50 simulations in particular, have been compared against observations. In particular, \citet{truong20} model the X-ray luminosity of TNG galaxies with an APEC-based approach (excluding star-forming gas, our $b=0$ case, see Appendix C of that work). Focusing on small apertures of $< R_{\rm e}$ for $0.3-5$\,keV, we compared against data from MASSIVE and ATLAS$^{\rm 3D}$ \citep{goulding16}, as well as the ETG sample of \citet{lakhchaura19}. Considering out to five times the effective radii ($< 5 R_{\rm e}$) for $0.3-6$\,keV, we compared against early-type data from \citet{babyk18}. In the galaxy mass range of interest here, the TNG50 simulations are well within the observational scatter of these datasets. Moreover, the key finding of \citet{truong20} was the prediction that, at fixed mass, star-forming galaxies should have X-ray brighter atmospheres than quenched galaxies \citep{truong21}, a result which also holds in the EAGLE simulation \citep{oppenheimer20} as well as in observations \citep[e.g][]{su15}. This suggests that the inner density structure of the \oviir emitting CGM, and so the importance of resonant scattering, is sensitive to galactic feedback processes including AGN-driven outflows, as already shown in Figure \ref{fig_enhancement_vs_mass}.

We conclude with a final comment on the impact of emission from the hot ISM itself. As a reminder, we incorporate a model for this component by treating star-forming gas as a two-phase medium \citep{spr03} and adopting the hot-phase fractional density and temperature for the respective \textsc{Cloudy} computation. This gives us a non-vanishing, centrally concentrated emission which can represent both the hot ISM and the presence of an AGN. We scale the emissivity of this component by a free parameter $b=10^{-4}$. This fiducial value shows that the underlying physical assumptions of this emission component are rough at best, and/or that intrinsic absorption of \oviir in the dense ISM is non-negligible \citep[as expected at high column densities;][]{lehmer22,vladutescu23}. We note that the ISM component is a small contribution in general, and that the diffuse, non star-forming gas dominates the total luminosity budget of essentially all galaxies in our model.

Figure \ref{fig_bparam} shows the impact of changing the strength of the hot ISM emission component. We return to our single galaxy test case, the same halo as in Figures \ref{fig_single_maps} and \ref{fig_single_profile}, showing the radial surface brightness profile of \oviir emission (top panel) and enhancement factor due to scattering (bottom panel). We contrast the intrinsic emission (dotted lines) with the scattered, observable emission (solid lines). We vary the $b$ parameter from $b=0$ (star-forming gas has no emission; red), to $b=10^{-4}$ (fiducial choice; blue), $b=10^{-3}$ (green), and $b=10^{-2}$ (orange). Only in the latter cases does this emission component start to have a non-vanishing impact on the observable surface brightness profiles. However, we rule out such choices as leading to overly high galaxy luminosities. For our fiducial choice, the hot ISM component of our emission model has little importance. Instead, it is the density, temperature, and ionization structure of the diffuse gas, as directly resolved in the simulations and used for the emissivity calculations, which dominates our predictions for the observability of \oviir emission from the circumgalactic medium of galaxies.


\section{Summary and Conclusions} \label{sec_conclusions}

In the near future, high resolution X-ray imaging spectroscopy will advance our understanding of the physics of the hot, virialized gas within dark matter halos. To date it has remained observationally elusive, but the hot circumgalactic medium of galaxies encodes signatures of the rich -- albeit complex -- interface of structure formation, galaxy evolution, and astrophysics.

In this study we have focused exclusively on one of the promising emission lines of highly ionized oxygen ions: the \oviir transition of the \ion{O}{VII} He-like triplet, with an energy of $0.574\,$keV in the soft X-rays. Due to its resonant nature, intrinsically emitted \oviir photons which originate from the bright halo center -- e.g., the hot star-forming ISM, a central AGN, or the dense inner CGM -- can scatter in the extended halo gas. This effect will enhance the observable surface brightness of the \oviir emitting circumgalactic medium.

To explore this phenomenon, we extend our Monte Carlo radiative transfer method, originally designed to study Lyman-alpha \citep{byrohl21,byrohl22}, to also treat resonant metal-line transitions, including \oviir. Taking advantage of its ability to ray-trace photons through unstructured Voronoi tessellations of space, we apply it to a sample of several hundred galaxies with $10^{10} < M_\star / \rm{M}_\odot < 10^{11}$ at $z=0$ from the high-resolution galaxies from the TNG50 cosmological magnetohydrodynamical simulation \citep{nelson19a,pillepich19}. Our key findings are:

\begin{itemize}
\item Resonant scattering significantly enhances the observable surface brightness of \oviir emission in the extended CGM of galaxies. 
\item The boost to the observable surface brightness can be large, even an order of magnitude effect. We adopt the ratio of scattered (observable) to intrinsic \oviir surface brightness at the virial radius ($R_{\rm 200c}$) as our `enhancement factor.' This is largest for low-mass galaxies, and decreases with increasing stellar mass. At intermediate masses $M_\star = 10^{10.5}$\msun, the mean enhancement is a factor of \textit{one hundred}. This decreases to a factor of ten by $M_\star = 10^{11}$\msun.
\item This scattering-induced enhancement boosts the entire extended CGM outside of $\gtrsim$\,10 kpc, and is generally stronger with increasing distance away from the galaxy. The mean enhancement from each surface brightness map is much larger than the median, indicating that localized bright regions are preferentially boosted. For $M_\star = \{10^{10.5},10^{11}\}$\msun, the median surface brightness at the virial radius is enhanced by factor of $\sim$\,20 and $\sim$\,3, respectively.
\item The enhancement of \oviir surface brightness depends, at fixed stellar mass, on galaxy properties. Specifically, galaxies with higher star formation rates, higher supermassive black hole (AGN) bolometric luminosities, and central galaxy \oviir luminosities all have significantly larger enhancement factors. In contrast, there is no trend with halo mass at fixed stellar mass. The scatter in this respect is comparable to the overall mass trend, suggesting that an identifiable subset of the galaxy population will be the most promising targets for detecting scattering enhanced \oviir emission from the CGM.
\end{itemize}

We have concentrated on \oviir in large part because of its observability with the Line Emission Mapper (LEM) X-ray observatory concept \citep{kraft22}. While LEM could focus its primary CGM science survey on the non-resonant \ion{O}{VII}{\small (f)} transition, which at $z = 0.01$ is redshifted out of the Milky Way foreground emission, our results suggest that \oviir at slightly higher redshifts could be a compelling target. Namely, the CGM of galaxies at $z=0.03$ would also clear MW foregrounds, and the marginal $(1+z)^4 \sim 10$\% surface brightness dimming due to this distance increase would be far out-weighed by surface brightness enhancement factors of many, to tens, due to the impact of resonant scattering.\footnote{For comparison with \oviir, we have also evaluated the impact of resonant scattering on the Ly$\alpha$ line of \ion{O}{VIII} at $\lambda = 18.95$\AA\, by running our radiative transfer method on a number of TNG50 halos from our sample with $M_\star \simeq 10^{10.5}$\,M$_{\odot}$. As expected, scattering has a much weaker impact for \ion{O}{VIII}. Enhancements are typically at the percent, to tens of percent, level, and roughly an order of magnitude less than for \oviir. For the galaxy shown in Figures \ref{fig_single_maps} and \ref{fig_single_profile}, the peak scattering enhancement of the \ion{O}{VIII} Ly$\alpha$ surface brightness is $50$\%, as compared to a factor of ten for \oviir. Scattering of this line may be more important for higher mass halos.} At this redshift, the 30' LEM field of view corresponds to approximately one physical Mpc, which is more than sufficient to capture the entire CGM of even the most massive halos considered here. In particular, the 1\,Ms detectability level of $\sim 10^{33}$\sbunits for LEM (Figure \ref{fig_stacked_profiles}) intersects our predicted surface brightness profile at a projected distance of $\sim 150$\,kpc for high-mass $M_\star = 10^{11}$\msun galaxies. The detectable \oviir CGM would therefore fill the inner detector. This will also be advantageous given the hybrid design that includes higher energy resolution pixels in the center: LEM will not only detect and spatially map this hot gas, but also characterize its kinematics.

In general, resonantly scattered X-ray line emission encapsulates a rich set of physics. Our methodology and results establish that forward modeling of this emission must account for radiative transfer effects. Future work can study additional observable implications, for example: the spectral distortions introduced by scattering, and possible surface brightness fluctuation smoothing effects. Scattering impacts all upcoming X-ray imaging spectroscopy missions and concepts, including XRISM, HUBS, and ATHENA. Properly treating this complexity opens new opportunities, as observations of resonantly scattering X-rays probe hot gas kinematics across several interesting regimes: microscopic turbulence, bulk flows, mixing, shock fronts, accretion streams, and feedback-driven outflows, from the circumgalactic medium of galaxies to the intracluster medium of the most massive dark matter halos in the Universe.


\section*{Data Availability}

The IllustrisTNG simulations, including TNG50, are publicly available and accessible at \url{www.tng-project.org/data}, as described in \cite{nelson19a}. Data directly related to this publication is available on request from the corresponding author.

\section*{Acknowledgements}

DN and CB acknowledge funding from the Deutsche Forschungsgemeinschaft (DFG) through an Emmy Noether Research Group (grant number NE 2441/1-1). We also thank the Hector Fellow Academy for their funding support. This work was further supported by the Deutsche Forschungsgemeinschaft (DFG, German Research Foundation) under Germany's Excellence Strategy EXC 2181/1 - 390900948 (the Heidelberg STRUCTURES Excellence Cluster). The material is based upon work supported by NASA under award number 80GSFC21M0002. IK acknowledge support by the COMPLEX project from the European Research Council (ERC) under the European Union’s Horizon 2020 research and innovation program grant agreement ERC-2019-AdG 882679. AO thanks Kristen Garofali for useful discussions. NW was supported by a CIERA Postdoctoral Fellowship. SVZ acknowledges support by the DFG project nr. 415510302. The TNG50 simulation was run with compute time awarded by the Gauss Centre for Supercomputing (GCS) under GCS Large-Scale Project GCS-DWAR on the Hazel Hen supercomputer at the High Performance Computing Center Stuttgart (HLRS). Additional computations were carried out on the Vera machine of the Max Planck Institute for Astronomy (MPIA) operated by the Max Planck Computational Data Facility (MPCDF).

\bibliographystyle{mnras}
\bibliography{refs}

\end{document}